# CHEMICAL ABUNDANCES IN VIRGO SPIRAL GALAXIES II: EFFECTS OF CLUSTER ENVIRONMENT [1]

Evan D. Skillman

Astronomy Department, University of Minnesota

116 Church St., S. E., Minneapolis, MN 55455

skillman@zon.spa.umn.edu

Robert C. Kennicutt Jr.

Steward Observatory, University of Arizona, Tucson, AZ 85721

robk@as.arizona.edu

Gregory A. Shields

Astronomy Department, University of Texas, Austin, TX 78712

shields@astro.as.utexas.edu

and

Dennis Zaritsky

UCO/Lick Observatory and Board of Astronomy and Astrophysics

University of California, Santa Cruz, CA 95064

dennis@lick.ucsc.edu

[1]Based on observations obtained at the Multiple Mirror Telescope, a joint facility of the University of Arizona and the Smithsonian Institution.

## ABSTRACT

We present new measurements of chemical abundances in H II regions in spiral galaxies of the Virgo cluster and a comparison of Virgo galaxies and field spirals. With these new data there now exist nine Virgo spirals with abundance measurements for at least four H II regions. Our sample of Virgo galaxies ranges from H I deficient objects near the core of the cluster to galaxies with normal H I properties, far from the cluster core. We investigate the relationship between H I disk characteristics and chemical abundances to determine whether dynamical processes that remove gas from the disk, such as ram pressure stripping by the intracluster medium, also affect the chemical abundances.

We divide the nine Virgo spirals into three groups of three galaxies each: those with strong H I deficiencies, intermediate cases, and those with no H I deficiencies. The three most H I deficient Virgo spirals have larger mean abundances (0.3 to 0.5 dex in O/H) than the spirals on the periphery of the cluster. This suggests that dynamical processes associated with the cluster environment are more important than cluster membership in determining the current chemical properties of spiral galaxies. There is also weak evidence of shallower abundance gradients in the H I deficient Virgo spirals.

We also compare the abundance properties of our Virgo sample to a large sample of field spirals studied by Zaritsky, Kennicutt, & Huchra (1994). Those authors found that the mean abundance of the disk gas increases with increasing maximum circular velocity, increasing luminosity, and decreasing (earlier) Hubble type (but with a large dispersion in the mean abundances and abundance gradients). The dispersion in the properties of field galaxies and the small size of the Virgo sample make it difficult to draw definitive conclusions about any systematic difference between the field and Virgo spirals.

Nevertheless, the H I deficient Virgo galaxies have larger mean abundances than field galaxies of comparable luminosity or Hubble type, while the spirals at the periphery of the cluster are indistinguishable from the field galaxies.

Simple, illustrative chemical evolution models with infall of metal-poor gas are constructed and compared to models in which the infall is terminated. The models are constrained by comparison with observed gas mass fractions, current star formation rates, and gas consumption times. The model results indicate that the curtailment of infall of metal-poor gas onto cluster core spirals may explain part of the enhanced abundance. However, additional work is needed, particularly modeling of the effects of truncating the outer gaseous disk within the context of models with radial gas transport.

The increased abundances in cluster core late-type spirals, relative to field galaxies, may be important in the interpretation of observations of these galaxies. Specifically, we point to possible effects on the Tully-Fisher and Cepheid variable distance determinations and the interpretation of colors in the Butcher-Oemler effect.

## 1. INTRODUCTION

Spiral galaxies are modified by the environment within a rich cluster of galaxies (see reviews by Haynes 1990 and Whitmore 1990, and the introduction to Henry *et al.* 1992). Such effects as tidal stripping, ram pressure stripping, and ablation by the intracluster medium must affect galaxy evolution. The cluster environment clearly influences galaxy morphology (Dressler 1980; Postman & Geller 1984), and so a natural question is whether it also affects the star formation history and chemical evolution of galaxies. Does the cluster environment predetermine galaxy morphology by solely influencing initial conditions or does it continually influence the evolution of a member galaxy? Comparing the chemical abundances of field and cluster members may provide clues to these fundamental questions. Moreover, a systematic abundance differential between cluster and field galaxies would have implications for the stellar populations and could impact a number of currently popular distance indicators.

The Virgo cluster, due to its proximity, provides the logical starting point for the study of environmental effects on abundances in cluster spirals. The Virgo cluster is irregular, and in some respects this compromises the study of environmental effects. However, the extensive catalog of the positions, morphologies, and radial velocities of 1277 galaxies by Binggeli, Sandage, & Tammann (1985) allows a definition of the cluster's structure. In addition, the structure of the hot gas component, which has been revealed through sensitive mapping of the X-ray emission made possible by ROSAT (Böringer et al. 1994), is very similar to the galaxy distribution.

Substantial evidence points to differences between the spiral galaxies in the core of the Virgo cluster and field spiral galaxies. Davies & Lewis (1973) first pointed to a deficiency in H I gas in Virgo spirals when compared to field spirals. Giovanelli & Haynes (1985) showed that galaxies in the outer parts of the Virgo cluster (and eight other clusters) were less deficient in H I than the galaxies in the core of the cluster. From H I synthesis observations of Virgo cluster spirals, Warmels (1986) concluded the H I deficiencies could be attributed to a depletion of the H I at large galactocentric radii. This was supported by Cayatte et al. (1994, CKBG).

Because the star formation history of a galaxy is related to its gas content, it is logical to ask if the chemical abundances in the cluster spirals are different from those of field galaxies. Shields, Skillman, & Kennicutt (1991, SSK) presented spectrophotometry of H II regions in five Virgo spirals. The Virgo spirals appeared to be considerably more metal rich ($\sim$ 0.3 dex) than the typical galaxy in their field sample. SSK suggested that such a difference could result from differences in infall or radial inflow rates. Henry *et al.* (1992, HPLC) and Henry, Pagel, & Chincarini (1994, HPC) conducted intensive studies of two Virgo spirals and found the evidence of a Virgo/field differential less convincing.

Hence, the issue of chemical abundance anomalies in the Virgo cluster is unresolved. The existing data sets have been too small to address the most fundamental physical question, whether the chemical properties of the H I deficient galaxies in the core of the Virgo cluster are different from those of undisturbed galaxies, in Virgo or elsewhere. We have undertaken a larger investigation which is aimed at addressing these problems in four ways by: 1) enlarging the Virgo sample; 2) choosing galaxies within Virgo that span the range of observed H I properties; 3) using published data on the H I distributions within galaxies to examine the relationship between the H I distribution and chemical abundance; and 4) by comparing the properties of the Virgo galaxies to a much larger sample of field spirals than was previously available.

This study includes measurements of more H II regions in the strongly H I deficient galaxies NGC 4571 and NGC 4689 (for which SSK had only one H II region each) and in several spirals with moderate or no H I deficiency. Our work increases the number of Virgo spirals with abundance measurements of four or more H II regions to nine. For a comparison field sample we use the data for 39 field spirals recently published by Zaritsky, Kennicutt, & Huchra (1994, ZKH).

In this paper we describe the observations and discuss the implications for the questions raised above. In Section 2 we discuss the sample selection, new observations, data reduction procedures, and O/H abundance determinations. In Section 3 we discuss the abundance properties of the Virgo spirals, test for any dependence of abundance properties on cluster location or H I deficiency, and use the ZKH results to determine whether there are any systematic differences between the field and cluster galaxies. In Section 4 we characterize and quantify the abundance differential seen for the H I deficient spiral galaxies. In Section 5 we assess these results in the context of a simple model for environmental effects on chemical evolution, and in Section 6 we outline future observational projects, and discuss the implications of elevated abundances for other studies of cluster spirals.

## 2. OBSERVATIONS

### 2.1. Galaxy Selection

The primary objective of this program is to measure H II region abundances for a large number of Virgo cluster spirals that span a larger range of H I deficiency. Following SSK, we chose to restrict our sample to late-type spirals, those classified as type Sbc or later in the Revised Shapley-Ames Catalog (RSA; Sandage & Tammann 1981). This was done to limit the variation in abundances due to variations in galaxy type (Oey & Kennicutt 1993 OK; ZKH) and to take advantage of the brighter H II regions in late-type galaxies.

We compiled a list of spirals in this type range that were mapped in H I by Warmels (1988) or Cayatte et al. (1990). We then selected a subset of galaxies exhibiting the full range of H I properties, from galaxies clearly deficient in H I (NGC 4501, NGC 4571, NGC 4689), to intermediate objects (NGC 4254, NGC 4321, NGC 4654), and finally to those located in the outskirts of the cluster and exhibiting apparently normal H I distributions (NGC 4303, NGC 4651, NGC 4713). By limiting our interest to Sbc and Sc galaxies, our sample excludes the most H I deficient spirals closest to the Virgo core. These objects are early-type spirals (Sa–Sb) which generally contain much fainter H II regions than the late-type spirals (Kennicutt, Edgar, & Hodge 1989, OK). Their H II regions are too faint for this program; future observations of these objects will provide an important additional test for environmentally induced changes in the chemical evolution of spirals. Figure 1 shows a map of the positions of the 9 Virgo cluster spirals relative to the main structural features identified and discussed by Binggeli, Tammann, & Sandage (1987; see their Figure 1).

Table 1 summarizes the properties of the Virgo sample. Listed are the projected separation from M 87, radial velocity, RSA and RC3 types (de Vaucouleurs et al. 1991), luminosity, rotation velocity as determined from the H I rotation curve, optical diameter as defined in the RC3, and H I to optical diameter ratio, where the H I diameter is defined as the diameter at which the H I surface density falls below 1 $M_\odot$ pc$^{-2}$ (following Warmels 1986). Warmels has argued that the latter ratio is a good indicator of the degree to which H I has been removed from the disk. Following ZKH, we have adopted a distance of 16.8 Mpc for the Virgo cluster galaxies (Tully 1988), which is in good agreement with the recently measured Cepheid distance of 17.1 Mpc for M100 (Freedman et al. 1994). Figure 2 shows the variation of $D_H/D_O$ for a sample of field spirals compiled by Warmels (1986), plotted as a function of RC3 T-type; the error bars indicate the dispersion among field spirals of the same type. The corresponding values for the Virgo spirals in our sample are plotted as points and labeled. This plot demonstrates the large range in $D_H/D_O$ for a given T-type in the field sample, the general tendency for smaller H I diameters among Virgo spirals as a whole, and the variation in $D_H/D_O$ within the Virgo sample.

The radial H I surface profiles of the Virgo spirals are shown in Figure 3. The points represent the azimuthally averaged H I surface densities from Warmels (1988) and CKBG, and for NGC 4571 from van der Hulst et al. (1987), while the solid lines represent the mean distributions for field spirals of the same morphological type (from CKBG). CKBG demonstrate that the radial H I surface density distribution of a field spiral galaxy is determined by its morphological type, if it is normalized to the optical size of the galaxy. We have normalized the observed profiles of the Virgo spirals in the same manner, using the $D_O$ diameters from the RC3. Figure 3 confirms the variation in H I deficiency within the Virgo sample, especially at the extremes (e.g., compare NGC 4501 and NGC 4689 with NGC 4651). We have ordered the panels in Figure 3 in increasing $D_H/D_O$ relative to the mean for the appropriate T-type, with the result that the galaxies group into the 3 deficiency groups discussed above. This subdivision into three groups is in general agreement with an independent classification by CKBG.

## 2.2. Spectrophotometry

Spectrophotometry for 30 H II regions in the Virgo spirals was obtained in April 1991 and May 1992, using the Blue Channel Spectrograph on the Multiple Mirror Telescope (MMT). The H II regions were identified from H$\alpha$ images obtained by RCK using a focal reducer CCD imager on the Steward Observatory 2.3 m telescope. The images also provided precise coordinates and nuclear offsets for the spectrophotometric observations. Preference in selection was given to bright H II regions that sample the greatest possible radial extent within each galaxy.

The observing setup at the MMT was identical to that described in SSK. The spectrograph was used with the 2-channel, 1024 element intensified Reticon photon-counting detector. When used with a 300 gpm grating, 3600 Å longpass blocking filter, and $2'' \times 3''$ aperture it provided a wavelength coverage of 3600−6800 Å at approximately 10 Å (FWHM) resolution. Object and sky apertures were beamswitched every 300 sec and total integration times ranged from 600 to 2400 sec, depending on the brightness of the object and sky conditions. Objects were observed at high zenith angles to avoid problems with differential refraction. Observing conditions ranged from clear to thin clouds, and because of the variable conditions and small aperture sizes only relative line fluxes are reported here.

The spectra were calibrated using observations of standard stars from Filippenko & Greenstein (1984) and Massey et al. (1988), as well as the usual internal wavelength and flatfield lamp calibrations. Fluxes and equivalent widths for the principal spectral lines were then measured using the SPLOT package in IRAF[2]. We confirmed that the SPLOT measurements agreed with the automated measuring techniques of ZKH. Line strength uncertainties were determined from the r.m.s. deviations in the adjacent continuum. In some cases the "sky" aperture fell within the galaxy, so the the object spectrum, constructed by subtracting off the sky spectrum, is contaminated. This can artificially increase the emission line equivalent widths and the uncertainties in the line strengths, and this effect has not been accounted for in our error estimates. The Balmer lines were corrected for underlying stellar absorption with an assumed constant equivalent width of 2 Å (McCall, Rybski, & Shields 1985, MRS), and then the reddening was determined from the relative Balmer line strengths assuming the values for an electron temperature of $10^4$ K and an electron density of 100 $cm^{-3}$ (Brocklehurst 1971). All line strengths were then corrected for reddening using the Galactic reddening law compiled by Seaton (1979), as parameterized by Howarth (1983). The resultant line strengths and associated uncertainties are listed in

---

[2]IRAF is distributed by the National Optical Astronomy Observatories, which is operated by the Association of Universities for Research in Astronomy, Inc. (AURA) under cooperative agreement with the National Science Foundation.

Table 2.

### 2.3. Abundance Determinations

The new spectra of 30 H II regions reported here, when combined with previously measured spectra from SSK, MRS, HPLC, and HPC, provide data for a total of 70 H II regions in the 9 Virgo galaxies listed in Table 1. Because accurate measurements of the electron temperature are not available for the regions, we have used the oxygen excitation index $R_{23} \equiv ([O\ II] + [O\ III])/H\beta \equiv I(\lambda3727 + \lambda4959 + \lambda5007)/H\beta$ to estimate the nebular abundances (Edmunds & Pagel 1984).

To ensure consistency among the abundances measured for the various field galaxies and the ZKH field galaxy sample, we have applied the same empirical $R_{23}$ abundance calibration as used by ZKH for all 70 of the Virgo cluster H II regions. The ZKH calibration is an average of three calibrations by Edmunds & Pagel (1984), MRS, and Dopita & Evans (1986). Since we are mainly interested in differential comparisons of the abundance properties of different objects, the precise choice of the $R_{23}$ calibration is not critical, and as argued by ZKH, the use of three separate calibrations provides an explicit, although incomplete, estimate of the error in the derived oxygen abundances. Typically, these calibrations are based on directly measured electron temperatures in the low abundance regime and photoionization model fits for high abundances (low excitation). Presumably the uncertainties are greatest at the high abundance end (see discussions in OK, Henry 1993, and Shields & Kennicutt 1995), which applies to most of the Virgo cluster H II regions. In any case, the derived abundance is monotonic in $R_{23}$ over the excitation range of interest here, and it should be meaningful at the very least as a relative abundance sequencing parameter.

Table 3 lists the abundances derived in this way for the entire Virgo H II region sample.

The O/H abundances listed in Table 3 show two associated errors. The first error is the result of propagating the formal errors of the emission line fluxes into the empirical calibration of $R_{23}$ by ZKH. The second set of errors (in parentheses) reflect the uncertainty of the $R_{23}$ calibration adopted by ZKH. In most cases, the latter are significantly larger than the former. Following ZKH, we use these larger errors throughout our analysis, unless noted otherwise. The derived abundances are plotted as a function of galactocentric radius (normalized to the effective radius, following McCall 1982) in Figure 4. The ordering of the galaxies is identical to that used in Figure 3.

Following ZKH, we have fitted a linear relation in log O/H versus radius to the abundances of the individual H II regions to determine the "mean" value of O/H at a fiducial radius. Like ZKH, we use a fraction of the isophotal radius, 0.4 $R_O$ as the fiducial radius. ZKH found this preferable because it is less sensitive than other choices to various selection effects (see discussion in ZKH). The results from these fits are given in Table 4 and plotted in Figure 4. Hereafter, we will refer to the mean value as that taken at 0.4 $R_O$. The errors in the mean abundances were determined in the standard manner for a weighted linear least-squares fit, where the weighting was inversely proportional to the formal errors in $R_{23}$ (the first set of errors in Table 3) and normalized to give Chi-squared equal to one. This prevents the uncertainties in the $R_{23}$ calibration from propagating into the uncertainty in the the mean abundance, i.e., we are interested in the uncertainty in the abundance differences between galaxies and not the absolute uncertainty in the abundance scale. For calculating the gradients, the second set of errors in Table 3 (the larger ones) were used, since the uncertainty in the absolute calibration is relevant to the uncertainty in the magnitude of the gradient.

Note that the outermost H II region in NGC 4651 lies well above the trend established by the other H II regions. As a result, the calculated mean abundance is slightly higher and the abundance gradient is slightly shallower than would result from a determination based on the inner H II regions. At this time, we have no reason to drop this discrepant point from the fit or to choose a more complex fitting function. From a comparison of the results in SSK, HPLC, and HPC, it is quite possible for an error in the abundance of a single H II region of 0.3 dex, and an error of this size would bring the outermost H II region in NGC 4651 into line with the trend established in the inner galaxy. However, if the inflection in the abundance distribution in NGC 4651 is real, then this galaxy deserves a more detailed abundance study.

## 3. THE EFFECT OF H I DEPLETION ON THE ABUNDANCES IN VIRGO SPIRALS

### 3.1. General Trends in the Virgo Spirals

We begin our analysis with a qualitative discussion of the abundances in the Virgo galaxies, based on a comparison of Figures 2, 3, and 4. We focus on the question of whether the H I deficient Virgo galaxies have different abundance properties than the Virgo galaxies with normal H I disks. Figures 2 and 3 are generally consistent as to which galaxies are deficient in H I, although the $D_H/D_O$ parameter is a limited indicator. NGC 4501, NGC 4571, and NGC 4689 are clearly deficient, the latter two even at small radii. NGC 4303, NGC 4651, and NGC 4713 are outlying galaxies with normal H I disks. The remaining three galaxies are intermediate cases. Comparing the H I data with the abundance patterns in Figure 4, we note that the peripheral galaxies have strong radial gradients in O/H and reach rather low values, $12 + \log(O/H) \leq 8.7$, at the largest radii. In contrast, the three most H I deficient galaxies show high abundances with little evidence for radial gradients (but note that the radial range of H II regions is very limited in the H I deficient spirals).

This is further illustrated in Figure 5, which shows the characteristic O/H values and gradients (from Table 4) for the Virgo galaxies as a function of $D_H/D_O$. In the top panel of Figure 5, a trend of decreasing O/H with increasing $D_H/D_O$ is evident for the Virgo galaxies. The top half of Figure 5 shows that our grouping into three categories is somewhat arbitrary as all nine Virgo spirals display a continuum of decreasing mean O/H with decreasing H I deficiency.

The bottom panel of Figure 5 illustrates the relationship between oxygen abundance gradient and H I deficiency. Here we tentatively see weak evidence for a trend of stronger gradient with decreasing H I deficiency. The most notable exception is NGC 4651, but, as noted above, a single linear fit to the radial abundance trend in NGC 4651 is a poor fit. The evidence for a trend in abundance gradient trend is weak, in part, because we have conservatively included the uncertainty in the $R_{23}$ calibration in the calculation of the uncertainty in the gradient. However, the main problem with this comparison is that only a small number of H II regions sampling a very restricted range of radius were measured in the H I deficient Virgo spirals. Note that this comparison is further complicated because the range of radii probed in each galaxy is different and because the radial normalization, the isophotal radius, may also be affected by the cluster environment.

The number of Virgo galaxies available is obviously less than one would like, and the number of H II regions sampled per galaxy is less than satisfactory for all but perhaps two or three of the galaxies. It is possible that the trends shown by the Virgo galaxies in Figure 5 are simply statistical flukes. On the other hand, one cannot make these trends go away simply by discarding any one galaxy. The three H I deficient galaxies have significantly higher characteristic abundances and possibly smaller gradients than the three galaxies with normal H I properties.

### 3.2. Comparison of the Virgo and Field Spirals

Do Virgo galaxies have different abundance properties than comparable field galaxies? We have argued above that the H I deficient galaxies in the cluster core show higher abundances than the outlying galaxies. Studies of abundance patterns in field galaxies demonstrate that the mean abundances of disks are systematically correlated with galaxy type, luminosity, and circular velocity (e.g., Pagel & Edmunds 1981, MRS, Garnett & Shields 1987, Skillman, Kennicutt, & Hodge 1989, Vila-Costas & Edmunds 1992 (VCE), OK, ZKH), and it is important to check that the patterns in Figure 5 are not due to underlying variations in those properties. For example, many of the Virgo spirals are very luminous galaxies, and so they might be metal-rich objects solely on that basis, independent of cluster environment.

To address this issue, we have used the ZKH survey to compare the abundances of the Virgo galaxies to those of the field galaxies, as functions of absolute magnitude, maximum disk circular velocity, $V_C$, and morphological T-type. In Figure 6 we have added the Virgo spirals to the field spirals from Figure 10 of ZKH. We exclude strongly barred spirals (RC3 "B" type) from the comparison because bars have been shown to affect the gradient slope (Pagel et al. 1979, VCE, ZKH, Martin & Roy 1994, Friedli, Benz, & Kennicutt 1994). (Note that although none of the Virgo sample spirals are strongly barred, four are classified as transitional "X" types: NGC 4303, NGC 4321, NGC 4654, and NGC 4713.) In Figure 6a we plot the mean O/H and gradient as a function of $M_B$. The H I deficient Virgo galaxies have higher oxygen abundances than the field galaxies, as expected from the earlier discussion. This is seen again in Figure 6b where the mean O/H value is plotted against $V_C$ and in Figure 6c, where the galaxies are plotted as a function of T-type. *The outstanding impression rendered by Figure 6 is that the three* H I *deficient Virgo spirals are all at the top of the abundance distributions of galaxies with similar properties.*

This is supported by a simple statistical test. Each of the three H I deficient Virgo galaxies has the largest characteristic abundance for galaxies of its corresponding T-type (cf. Figure 6c). The 1/N probability of this happening for each galaxy, if the abundances are distributed randomly within galaxies of the same T-type, are 0.333, 0.083, and 0.111 for NGC 4501, NGC 4689, and NGC 4571, respectively. Because each of these presumably represents an independent event, the probability that all three would have the largest abundances within galaxies of their T-type by chance is 0.003. We discuss other possible factors in greater detail below, but we conclude here that there is strong evidence that the H I deficient galaxies have elevated characteristic abundances.

In making this comparison between Virgo spirals and field spirals, we need to be careful that we have not introduced a bias in the composition of the sample. For example, our simple statistical test could be biased by other factors that affect abundance, i.e., we know that luminosity and abundance are correlated. This is related not only to the question of the reality of any Virgo abundance differential, but also to the broader issue of how Hubble types are influenced by cluster membership. In Figure 7, we plot $M_B$ and $V_C$ versus T-type. There is no evidence that the H I deficient group suffers a bias in comparison to the field sample. Among the H I deficient Virgo spirals, NGC 4501 is near the top of the distribution in $M_B$ and $V_C$, but NGC 4571 and NGC 4689 are near to the middle of the distributions in $V_C$ and nearer the bottom in $M_B$. Thus, their higher mean abundances do not appear to be due to their intrinsic properties of luminosity or mass. However, the sub-sample of intermediate galaxies may be skewed. All three intermediate galaxies are near the top of the distributions in $M_B$ and $V_C$ for their T-types.

Finally, we checked our result from the preceding section showing a dependence of O/H on $D_H/D_O$ in the Virgo spirals; i.e., we needed to check for such a trend in field spirals. Constructing a comparison for the field galaxies is not straightforward. Only about two-thirds of the galaxies in the ZKH sample have published H I synthesis observations, and unlike the uniform H I observations of the Virgo cluster sample, the H I observations of the field galaxies were obtained with a large range in linear resolution. Despite such difficulties, a comparison was constructed, and we found no evidence of a trend in O/H with $D_H/D_O$ in the field spirals. This suggests that the trend is due to the effects of the cluster environment and not a characteristic of galaxies in general.

### 3.3. Are the H I Deficient Virgo Spirals Chemically Different?

From a comparison of five Virgo spirals and nine field spirals, SSK found the Virgo spirals to be considerably more metal rich ($\sim$ 0.3 dex) than comparable field spirals. HPLC added observations of NGC 4303 and HPC added observations of NGC 4254. These observations were found to be in general agreement with those of SSK. From a comparison of the three Virgo spirals with more than 10 observed H II regions with five field spirals, HPC concluded that the Virgo spirals might be more metal rich than the field spirals by about 0.15-0.20 dex. SSK, HPLC, and HPC all suggested that a uniform field sample and an understanding of effects due to galaxy mass and structure were needed before firm conclusions about environmental influences could be drawn.

We have addressed these concerns by increasing the number of observed Virgo galaxies to nine, increasing the number of observed H II regions in these galaxies, and making the comparison to the large, uniform sample of ZKH. By concentrating on the comparison between the H I *deficient* Virgo spirals and the field and peripheral cluster galaxies, we have clarified the evidence of an abundance differential. In retrospect, the need to separate the Virgo galaxies into H I deficient and H I normal classes is an obvious one. The intended experiment is to test for the effects of cluster environment on the ISM abundances of cluster spirals, so it makes sense to search for the effects in those galaxies which show evidence of altering by the cluster environment.

We conclude that H I deficient spirals in the cluster core have larger abundances than comparable field spirals or outlying Virgo spirals. It is noteworthy that the outlying Virgo spirals are indistinguishable from field galaxies of comparable type and luminosity. Quantifying the size of the abundance differential is problematic due to the small sample size, the dispersion in the field comparison sample, the change in both mean abundance and gradient, and the variation in differential with the degree of H I deficiency. We will return to this in the next section.

## 4. QUANTIFYING THE ABUNDANCE DIFFERENTIAL

To model the effect of cluster environment on spiral abundance patterns, it is necessary to quantify the size of the abundance differential that we claim to see. For this, we concentrate on the three most H I deficient galaxies in the Virgo sample. Taking the characteristic abundances grouped by morphological type and averaging the logarithmic values, we find an average characteristic abundance of $8.63 \pm 0.48$ for the 7 T = 6 galaxies from the ZKH sample. This is 0.61 dex (or 1.3 $\sigma$) lower than the characteristic value for NGC 4571. The 9 T = 4 galaxies yield an average of $9.05 \pm 0.13$, which is 0.23 dex (or 1.8 $\sigma$) lower than NGC 4689. The 2 T = 3 galaxies have an average of $9.05 \pm 0.03$, which is 0.27 dex (or 9 $\sigma$) lower than NGC 4501 (although with only 2 comparison galaxies, a statistical comparison for NGC 4501 does not seem justifiable).

The above analysis focuses on characteristic abundances, but it is useful to compare entire chemical abundance distributions to ensure that the choice of a fiducial radius is not biasing our result. For example, by restricting our comparison to the characteristic abundances it is possible that we could underestimate the chemical composition difference in the H I deficient Virgo spirals (since the observed shallower gradients means that the effect is largest at the last measured point). To investigate this possibility, we have constructed Figure 8, where the radial abundance patterns of the three most H I deficient Virgo spirals are compared to the abundance patterns of the non-barred field spiral galaxies of comparable morphological type from the ZKH compilation. The most striking case is NGC 4571, shown in the top panel of Figure 8. NGC 4571 has measured H II regions out to $R_E \approx 1.3$ and for which the data are consistent with no gradient. Comparing the O/H at the last measured point for NGC 4571 with the median value for the comparison field galaxies, the abundance differential is, again, roughly 0.6 dex. The comparison for NGC 4689 is less dramatic. A similar comparison of the O/H at the last measured point to the median for the comparison sample yields a differential of roughly 0.2 dex. Note, however, that NGC 4869 has a lower luminosity than all but one of the comparison field galaxies, and should, therefore, be expected to have *lower* abundances than the comparison galaxies. NGC 4501 has a small comparison sample, but shows a differential similar to that seen in NGC 4689. Thus, using characteristic abundances and outermost measured points give similar answers.

Finally, in Figure 9, we construct a comparison of O/H versus $R/R_E$, disk V-band surface brightness, and gas mass fraction ($\mu$), similar to Figure 1 of SSK. We plot all of the individual H II regions in the Virgo sample using different symbols for the H II regions in H I deficient, intermediate, and H I normal galaxies. This comparison is relatively free of distance uncertainties. The disk V-band surface brightnesses were taken from the radial averages of Kodaira et al. (1986). The gas mass fractions were calculated by taking the radial average molecular gas masses from Kenney & Young (1989), the radial average H I gas masses of Warmels (1986) and assuming a V-band mass-to-light ratio of 3 for the disks. The gas mass fractions must be considered very uncertain in an absolute sense (particularly due to the uncertainty of converting CO observations to a molecular mass surface density and the assumption of a constant mass-to-light ratio for the stellar disk), but there should be, at least, some sense in the relative ranking of radial gas mass fraction. As pointed out by SSK, the value of plotting O/H versus gas mass fraction is that it shows that the elevated abundances in the H I deficient galaxies are not a result of otherwise normal galaxies simply converting most of their gas into stars.

In Figure 9 we find that the H II regions from the H I deficient galaxies form a relatively distinct group at the highest abundances in all three comparisons. The offset between the H I deficient galaxies and the H I normal galaxies appears to be of $\sim$ 0.3 to 0.5 dex in the top two panels. The middle panel, the comparison with disk V-band surface brightness, appears to show the greatest order in the separation between the H I deficient, intermediate, and H I normal groups. VCE have shown that local metallicity correlates well with radially averaged disk surface brightness, and have argued that the correlation is improved when total surface brightness (disk plus bulge light) is used. Ryder (1995) has confirmed the metallicity – surface brightness relationship using I-band surface photometry for 6 southern spirals and F-band surface photometry for 6 of the spirals in the ZKH sample. Our results show that the abundance differential appears prominently in the comparison with surface brightness. In the comparison with $\mu$, there is less order in the distributions. This could be interpreted as indicating that the different galaxies have experienced different evolutionary histories, or may just reflect the uncertainty of the calculation of $\mu$.

It is important to recall that the most H I deficient galaxies in the Virgo cluster are all early type spirals, which we excluded from our sample due to the difficulty of measuring the abundances of their very faint H II regions (and additionally, due to the less extensive field galaxy comparison sample). Conservatively, we conclude from these data that abundance differentials at least as large as 0.2 and plausibly as large as 0.5 dex are produced by the cluster environment. When plotted in the metallicity – surface brightness plane, all three H I deficient galaxies stand apart from the H I normal spirals.

## 5. INTERPRETATION OF THE VIRGO ABUNDANCE DIFFERENTIAL

Having empirically established the abundance differential and provisionally quantified the magnitude of the effect, it is reasonable to ask if the observed effect is consistent with the physical processes known to be operating in the Virgo cluster environment. Our framework is that spiral galaxies are falling into the Virgo cluster (Tully & Shaya 1984), and that removal of gas from the infalling spirals by the cluster environment is primarily responsible for the H I deficiencies observed (Warmels 1986). We review the evidence supporting these mechanisms and then turn to the question of whether the observed abundance differential is consistent with this scenario.

### 5.1. Overview of Cluster Processes

Tully & Shaya (1984) studied the phase space distribution of the galaxies in and near the Virgo cluster and developed a mass distribution model in which galaxies within about 8 Mpc of the Virgo cluster are now falling back toward the cluster. Two of our H I normal galaxies (NGC 4303 and NGC 4713) are predicted to enter the Virgo cluster environment within the next 0.5 Gyr. Within their model, the majority of the elliptical galaxies in Virgo have been there for a substantial fraction of the age of the universe, while most spirals have entered the cluster in the last one-third to one-half of the age of the universe. This implies that the Virgo spirals were formed in lower density environments, more like field galaxies, and have only lately entered the high density cluster environment. Additionally, Dressler (1986) compared the velocity dispersions of H I deficient and normal spirals in nine clusters (including Virgo) and found that the H I deficient galaxies were more likely on radial orbits that would carry them into the cluster cores. These results correspond well with our observation that the spirals in the periphery of the cluster (the H I normal galaxies) have abundance patterns similar to those observed in field spirals and that the interaction with the cluster environment, not simply cluster membership, is the determining factor for H I deficiency.

Through H I synthesis mapping of 36 Virgo spiral galaxies and a comparison with a compilation of 63 field spirals, Warmels (1986) confirmed the H I deficiencies of the central Virgo cluster galaxies. He identified ram pressure stripping of the outer neutral gas by the intracluster medium as the main cause of H I deficiency based on four lines of evidence. First, the $D_H/D_O$ ratios for the spirals in the inner three degrees of the cluster are much smaller than for both the cluster spirals, which are at projected radii greater than five degrees and the field spirals. Second, the scale lengths of the declining part of the radial H I distributions of the inner Virgo spirals are smaller than those of the outer Virgo spirals. Third, only the H I surface densities in the outer parts of the H I deficient galaxies appear to be affected (although this conclusion has been questioned by later studies, Kenney & Young 1989; CKBG). Fourth, the H I distributions of the inner Virgo galaxies show large asymmetries. Warmels compared his observations with the predictions from models of galaxy collisions, tidal stripping, thermal evaporation, and ram-pressure stripping, and concluded that the observations were only consistent with ram-pressure stripping models. Following the simple model of Gunn & Gott (1972), and assuming an intracluster density of $5 \times 10^{-4}$ cm$^{-3}$ and a galaxy velocity of 500 km s$^{-1}$, Warmels was able to reproduce the size of the effects observed. Subsequently, this simple model was shown to work for the Virgo dwarf galaxy IC 3365 (Skillman et al. 1987), and Sancisi, Thonnard, & Ekers (1987) discovered a cloud of H I which appeared to be stripped completely from a dwarf galaxy in the vicinity of NGC 4472 (see also Henning, Sancisi, & McNamara 1993).

Kenney & Young (1989) observed CO emission from 40 Virgo cluster spirals, and compared the radially averaged neutral atomic and molecular gas distributions. They concurred with Warmels' conclusion that the outer parts of the inner Virgo spirals were H I deficient, but noted that the inner parts of those galaxies were also H I deficient (by a factors of $\sim 2 - 3$). Perhaps surprisingly, Kenney & Young found that the CO emission from the H I deficient Virgo galaxies was *not* deficient. This is especially interesting in light of the results of Kennicutt (1983), who found that star formation rates, inferred from H$\alpha$ luminosities, are lower in spirals in the Virgo cluster core than in comparable field galaxies. However, Kennicutt (1989) has shown that star formation rates do not correlate well with the amount of molecular hydrogen present in the disk as inferred from CO observations. Because the CO emission is confined to the centers of spiral galaxies, the normal CO emission from H I deficient spirals is not in conflict with a model of ram pressure stripping of the H I deficient cluster spirals. Note also, that the above discussion assumes a metallicity independent conversion of CO emission to molecular gas content.

CKBG have divided the Virgo cluster galaxies into four groups: (I) normal, (II) stripped by transport processes, (such as turbulent viscosity and thermal conduction, Nulsen 1982), (III) ram-pressure stripped, and (IV) completely stripped (but recovering). The observational distinction between groups II and III is that group III galaxies have smaller projected cluster radii, larger central H I surface densities, and smaller $D_H/D_O$ ratios than group II galaxies. CKBG conclude that the group II galaxies are too far from the cluster core for ram-pressure stripping to dominate and that transport processes, which would be effective over the entire disk, dominate. Of our nine galaxies, all three galaxies in our H I deficient group and the intermediate case NGC 4321 are classified as group II, while the rest of the sample are classified as group I galaxies. Our preference for late-type spirals has excluded the type III and IV galaxies, as all are T-type 2 and 3. If the reasoning presented by CKBG is true, then we should be trying to interpret the abundance differential between our group I and group II galaxies within the framework of a stripping model based on turbulent viscosity and thermal conduction and not ram-pressure stripping.

### 5.2. Simple Models with Termination of Infall

SSK argued that galaxies immersed in the hot gas of the Virgo cluster core would not accrete primordial gas from their surroundings. The lack of accretion might cause these galaxies to evolve chemically like the "simple" closed box model of chemical evolution, whereas field galaxies would have their abundances depressed by infall of metal poor gas. The simple model predicts the relation $Z = y \ln \mu^{-1}$, where $Z$ is the abundance of heavy elements, $y$ is the yield of heavy elements per unit star formation, and $\mu$ is the gas mass fraction. SSK's Figure 3 suggests that the curtailment of infall might elevate the abundances in the Virgo spirals by as much as ~0.2 dex. Since the H I deficient Virgo spiral H II regions have metallicities from 0.2 dex to 0.6 dex higher than the normal Virgo spiral H II regions, it would appear that curtailment of infall alone is insufficient to produce the observed abundance differential. We will not attempt an exhaustive discussion of alternative explanations for the Virgo abundance differential, but as an illustration, we have constructed some simple models to obtain rough estimates of the differential effects on abundance, color, equivalent width, and gas consumption time that might result from termination of infall of metal poor gas.

Three observational constraints will be used. The first constraint is the gas mass fraction, $\mu$. While the simple closed box model predicts $Z = y \ln \mu^{-1}$; an infall model with instantaneous recycling and constant $M_g$ predicts $Z/y = 1 - \ exp(1 - \mu^{-1})$ (Larson 1972). The second constraint is the present star formation rate (SFR). This is parameterized by the ratio of the present SFR to the average SFR over the history of the disk, $b$ = SFR $(t)/$<SFR> (Scalo 1986). The conversion of $b$ values into the observable properties of H$\alpha$ equivalent widths, EW (H$\alpha$), and B–V colors is very model dependent. However, Kennicutt, Tamblyn, & Congdon (1994, KTC) give relationships between $b$ and the B–V colors and EW (H$\alpha$) for models with steady or decreasing star formation rates similar to our models and we use them here. The third constraint is the gas consumption time or "Roberts" (1963) time, $\tau_R$ (see discussion in KTC). Following KTC, we distinguish between two values of the gas consumption time: $\tau_R$, the total gas mass divided by the SFR; and $\tau_{R,d}$, the corresponding ratio within the optical disk ($R_O$) alone.

In Table 5, we list representative values for the Virgo galaxies in this study and for the nonpeculiar field galaxies in Table 5 of KTC. The first line presents the average values of the oxygen abundances for each of the three Virgo sample subgroups. The EW (H$\alpha$) and B–V values for the Virgo galaxies are taken from Kennicutt & Kent (1983) and the RC3 respectively. The EW (H$\alpha$) and B–V values for the field Sc's are taken from Kennicutt (1983), who also finds that the ratio of hydrogen to blue luminosity, $M_H/L_B$, is about 0.26 dex smaller in Virgo Sc's than those in the field. The three Virgo groups show a progression in EW from 10 Å for the H I deficient galaxies to 37 Å for the normal galaxies, consistent with the proposition that the outlying Virgo galaxies are like normal field galaxies. The progression in colors likewise resembles the difference between Virgo and field Sc's quoted by Kennicutt (1983). From Table 1 of KTC, average values of $b$ are 0.19 for the Virgo core galaxies and 0.56 for the field galaxies.

Roberts time values are also taken from KTC. For our sample, the values of $\tau_R$ and especially $\tau_{R,d}$ are smaller for the normal galaxies than for the H I deficient galaxies. This means that the SFR in the optical disk is depressed by more than the gas mass, and one might speculate that the SFR drops precipitously as the gas surface density falls toward a threshold value (Quirk 1972; Kennicutt 1989). We adopt a local gas consumption time of 2.5 Gyr as the baseline (field spiral), the same as in the KTC model and roughly consistent with the value of $\tau_{R,d} = 2.7$ Gyr for the whole KTC sample.

We construct models that represent a typical field Sc galaxy, and then examine the changes in chemical abundance and other observables when the infall is terminated. Our models follow the evolution of the abundance of a "primary" element that is produced by short-lived stars and ejected effectively instantaneously into the interstellar medium. The abundance is calculated in terms of the *gross* yield, $y_g$, the mass of the heavy elements produced per unit mass of gross star formation, undiminished by return of gas by evolving stars. Note that the above expressions for the "simple" and Larson models involve the *net* yield; for instantaneous recycling this is $y = y_g/(1-R)$, where R is the fraction of the stellar mass returned to the interstellar medium. Throughout, we take $y_g = 1$ (i.e., $Z \;=\; Z/y_g$). Only a single zone is considered, with an initial gas mass taken to be unity. There is no escape of gas or radial flow. The infall rate is parametrized as $f(t) = f_0 \; exp \; (-t/\tau_f)$, where mass is measured in the same arbitrary, dimensionless units as $M_g$, and $t$ is in units of Gyr. The star formation rate, SFR $(t)$, is given by one of two alternative prescriptions: (1) a linear Schmidt (1963) law, SFR $(t) = M_g/\tau_s$; or (2) SFR $(t) = f(t) + r(t)$, where $r$ is the rate at which gas is returned by dying stars. The latter gives a constant gas mass, $M_g(t) = 1$ (Larson 1972), which is motivated by the idea of a surface density threshold for star formation. Mass return from stars of various ages and corresponding metallicities were computed using the return rates given in Figure 1 of KTC for the K83 IMF. All models have final ages of 10 Gyr, and infall ceases at 7 Gyr in the truncation models.

Table 6 lists the parameters and results of some illustrative models. The models are labeled according to S or T for Schmidt Law or Threshold, H or L for high or low infall rate, and N or C for no cut-off or cut-off of infall. The first four models employ a SFR proportional to gas density with $\tau_R = \tau_s = 2.5$ Gyr. To limit the number of free parameters, we have taken a constant infall rate $f(t) = f_0$. The value of $b$ is then tuned by adjusting $f_0$. Model SHN has $f_0 = 0.2$, so that infall brings in 20 percent of the initial gas mass in 1 Gyr. This model has $b = 1.02$ after 10 Gyr. However, the adopted IMF returns 46 percent of the mass of stars formed in an instantaneous burst by an age of 10 Gyr, and most of this amount within 1 Gyr. Thus, we have $R \approx 0.46$, and the effective timescale of conversion of gas to long lived stars is $\tau_{s,net} \approx \tau_s/(1-R) \approx 4.6$ Gyr. Hence, infall roughly balances the net star formation, keeping $M_g$ and the SFR roughly constant. This model reaches an abundance $Z = 1.49$, and a gas fraction $\mu = 0.30$. For comparison, the simple model predicts $Z = 2.02$, and a Larson infall model would predict $Z = 1.64$, in better agreement with our model (for the simple and Larson models we have taken the net yield to be $y = y_g \; M_{s,gross}/M_{s,net}$ evaluated at the present, i.e., $t = 10$ Gyr).

Model SHC shows the result of abruptly cutting off infall 3 Gyr ago in a model otherwise identical to Model SHN. The result is an increase in Z of 0.19 dex, which is a substantial fraction of the Virgo abundance differential indicated by our observations. The star formation parameter drops from $b = 1.02$ to 0.61 and $\mu$ decreases from 0.3 to 0.2. From the KTC models, this increase in $b$ corresponds to an increase in B–V of about 0.08 dex, in rough agreement with the Virgo/field differential, but the drop in EW (H$\alpha$) by about 0.14 dex is about half that observed.

Models SLN and SLC give a similar comparison for an infall rate decreased by a factor of four. Model SLN, with no cutoff of infall, results in lower values of $b$ and $\mu$ and a higher value of $Z$ than in SHN. Cutting off infall again decreases $b$, $\mu$, and increases $Z$. The drop in $b$ again implies roughly the observed increase in B–V, but somewhat too little change in EW (H$\alpha$). As expected, when infall is less important, the increase in Z is less (0.12 dex). A larger increase in Z would require a larger $(t - t_c)/\tau_s$ ratio so that ln $\mu$ would decrease more. We conclude from this set of models that a cutoff of infall several billion years ago is roughly consistent with the colors of the Virgo spirals versus the field, but can account for only part of the observed abundance differential.

The next four models in Table 6 are analogous pairs of models for the case of a constant gas mass, $M_g = 1.0$. Note that, in contrast to the previous models, when infall is cut off, the SFR drops precipitously, because only recycled gas is available for new star formation.

As a result, the infall cut-off models result in higher values of $\mu$. For the first pair (TLN and TLC), the model with infall cut-off shows no increase in Z, too large a value of $\tau_R$, and too small a value of $b$. Increasing the infall rate (models THN and THC) results in a Z increase comparable to the Schmidt Law models, but the problems with $\tau_R$ and $b$ persist. Increasing the infall rate further in the cut-off models results in higher Z and lower $\tau_R$, but $b$ remains low.

We conclude from this discussion that, if late type field spirals experience continuing infall, then curtailment of this infall when a galaxy enters the cluster environment may elevate its chemical abundances by an amount of order 0.1 to 0.2 dex. There will be a corresponding reduction in the SFR, roughly consistent with colors and H$\alpha$ equivalent widths. This conclusion applies to models in which star formation continues after infall cutoff in rough proportion to the gas mass. However, models with a constant gas mass fail to produce significant abundance differentials, when constrained by colors and star formation rates.

## 6. FUTURE WORK AND IMPLICATIONS FOR OTHER STUDIES

The simple models described in the last section are not able to directly address the effects of the cluster environment on the radial abundance gradients. All of the models with continuing infall of significant magnitude have abundances close to those predicted by the analytic expression for the Larson infall model, essentially $Z \approx y$ for small $\mu$. Such models should result in very weak radial abundance gradients, for radii at which $\mu$ remains small. In contrast, strong radial gradients in O/H and other ratios are typically observed (e.g., Figure 8). This is one reason to consider models in which strong radial gas flows occur. A number of authors have considered the role of radial gas flows in the chemical evolution of disk galaxies. Mayor & Vigroux (1981) argue that infall will induce radial flows, and derive a radial abundance gradient similar to that observed in the Milky Way. Radial gradients caused by radial inflow are discussed by Tinsley & Larson (1978) and Götz & Köppen (1992). Lacey & Fall (1985), Clarke (1989), Sommer-Larson & Yoshii (1989, 1990), and Edmunds & Greenhow (1995) have explored models in which gas can flow both inward and outward.

SSK noted that the removal of the outer H I disk would prevent radial inflow of metal poor gas. The lack of inflow might lead to flat abundance gradients and elevated overall abundance values. Our conclusion that curtailment of infall contributes at most a fraction of the observed Virgo abundance differential emphasizes that other processes, including radial inflow, should be considered. We hope that the simple modeling in the previous section will help to illustrate how the Virgo abundance differential, together with other data such as colors, star formation rates, and gas masses, can be used to gain insight into the chemical evolution of galaxies.

Determining the mechanism that gives rise to the abundance differential is beyond the scope of this investigation, and clearly there remain some potentially fruitful observational projects. First, it would be desirable to extend this type of study to the earlier type spirals in the Virgo cluster core, including the group III and IV galaxies of CKBG. The intriguing proposition that the inner disks of the group III galaxies are unaffected suggests that one might not see an abundance differential there. The hypothesis that group IV galaxies have an ISM which consists purely of recent stellar mass loss may afford the opportunity for direct measurement of stellar yields. Additional observations of H II regions in the galaxies studied here would not only improve the statistical arguments, but might provide additional information beyond the characteristic numbers discovered here (see Zaritsky 1992). It would also be of great interest to carry out a similar investigation for other clusters. In rich clusters like A262, A1367, and Coma, ram-pressure stripping will be important out to a greater radius, and large samples of severely truncated (group III) galaxies are present (Giovanelli & Haynes 1985). Finally, observations of the abundances in the spiral galaxies in the Hydra I cluster would provide a valuable control sample, as these cluster spirals show no evidence of a correlation of H I deficiency with cluster radius (McMahon 1993).

This work has implications beyond a better understanding of the chemical evolution of field and cluster spiral galaxies. Foremost are implications for distance determination studies. The elevated abundances of H I deficient cluster spirals may impact both Tully-Fisher and Cepheid variable distance determinations. For example, if the dust content of spiral galaxies scales with abundance (van den Bergh & Pierce 1990), and the opacity corrections for galaxy luminosity are important (see Giovanelli et al. 1994 for a recent discussion), then there will be a systematic offset between cluster and field spiral galaxy Tully-Fisher relationships. It is also possible that chemical abundance studies will provide insight into the processes by which gas is removed from spiral galaxies as they enter the cluster environment. This may, in turn, lead to a more secure interpretation of current studies of the evolution of cluster galaxies (e.g, Dressler et al. 1994). A better understanding of the chemical abundance patterns in both field and cluster spirals would put these studies on firmer ground.

Throughout this work we have benefited from discussions with M. Edmunds, D. Garnett, R. Henry, B. Pagel, J. van Gorkom. We gratefully acknowledge support from NASA-LTSARP grant NAGW-3189 (EDS), NSF grants AST-9019150 and AST-9421145 (RCK), and STScI grant GO-4276.03-92A (GAS). RCK would like to thank the Department of Astronomy at the University of Texas for their generous hospitality and the support of the Beatrice M. Tinsley Centennial Professorship. GAS would like to acknowledge support from the William F. Marlar Memorial Foundation, and the hospitality and support of the Department of Space Physics and Astronomy, Rice University, during Spring 1995. We all acknowledge the capable MMT operators Carol Heller, John McAfee, and Janet Miller. We thank an anonymous referee for very helpful comments.

TABLE 1

VIRGO CLUSTER SPIRAL GALAXY PROPERTIES

| Name | $R_{M87}$ [a] | Velocity[b] | Type[c] | T-Type[d] | $L_B$ [e] | $V_c$ [f] | $D_O$ [g] | $D_H/D_O$ [h] | $R_E$ [i] |
|---|---|---|---|---|---|---|---|---|---|
| NGC 4254 | 3.7° | 2354 | Sc(s) | 5 | 42 | 307 | 5.6′ | 1.37 | 0.87′ |
| NGC 4303 | 8.2° | 1486 | Sc(s) | 4 | 43 | 216 | 6.5′ | 1.32 | 1.11′ |
| NGC 4321 | 4.0° | 1540 | Sc(s) | 4 | 47 | 201 | 7.6′ | 1.08 | 1.39′ |
| NGC 4501 | 2.1° | 2233 | Sbc(s) | 3 | 52 | 278 | 7.1′ | 0.98 | 1.39′ |
| NGC 4571 | 2.4° | 298 | Sc(s) | 6.5 | 9 | 165 | 3.7′ | 1.02 | 0.87′ |
| NGC 4651 | 5.2° | 772 | SBc(s) | 5 | 18 | 250 | 4.1′ | 2.08 | 0.69′ |
| NGC 4654 | 3.5° | 992 | SBc(rs) | 6 | 23 | 198 | 4.9′ | 1.50 | 1.11′ |
| NGC 4689 | 4.6° | 1578 | Sc(s) | 4 | 13 | 185 | 4.4′ | 0.80 | 1.11′ |
| NGC 4713 | 8.6° | 588 | Sc(s) | 7 | 8 | 137 | 2.7′ | 2.07 | 0.44′ |

[a] Projected distance from M87.

[b] Radial velocity with respect to the Galactic center of rest from the RC3.

[c] Morphological type from the RSA.

[d] Morphological type from the RC3.

[e] Blue Lumminosity in units of $10^9$ $L_\odot$ assuming all galaxies at the distance of 16.8 Mpc and using $B_T^o$ from the RC3.

[f] Maximum rotation curve velocity in km $s^{-1}$ from Warmels (1986).

[g] Corrected optical diameter $D_O$ from the RC3.

[h] HI to optical diameter ratio as defined by Warmels (1986).

[i] Disk effective radius derived from the photometry of Kodaira et al. (1986).

TABLE 2a
Dereddened Spectra

| λ Identification | NGC 4501 +018 +012 | NGC 4501 −043 −005 | NGC 4501 −027 +062 | NGC 4501 −094 +049 | NGC 4501 −068 +093 | NGC 4571 −017 −062 | NGC 4571 +055 +018 | NGC 4571 −029 −011 |
|---|---|---|---|---|---|---|---|---|
| 3727 [O II] | 0.22 ±0.07 | 1.00 ±0.12 | 0.27 ±0.07 | 1.83 ±0.14 | 0.63 ±0.08 | 0.58 ±0.07 | 1.07 ±0.12 | 0.46 ±0.29 |
| 4101 Hδ | ··· | 0.34 ±0.10 | ··· | 0.36 ±0.05 | 0.26 ±0.13 | 0.22 ±0.06 | ··· | ··· |
| 4340 Hγ | 0.45 ±0.06 | 0.42 ±0.07 | 0.50 ±0.10 | 0.51 ±0.05 | 0.47 ±0.06 | 0.40 ±0.04 | 0.43 ±0.07 | 0.48 ±0.24 |
| 4861 Hβ | 1.00 ±0.07 | 1.00 ±0.07 | 1.00 ±0.07 | 1.00 ±0.06 | 1.00 ±0.07 | 1.00 ±0.05 | 1.00 ±0.08 | 1.00 ±0.14 |
| 4959 [O III] | ··· | ··· | ··· | ··· | ··· | ··· | ··· | ··· |
| 5007 [O III] | ≤ 0.09 | ≤ 0.09 | ≤ 0.08 | 0.13 ±0.04 | ≤ 0.06 | 0.04 ±0.02 | 0.12 ±0.07 | 0.12 ±0.08 |
| 6563 Hα | 2.86 ±0.14 | 2.86 ±0.14 | 2.86 ±0.14 | 2.86 ±0.14 | 2.86 ±0.14 | 2.86 ±0.15 | 2.86 ±0.16 | 2.86 ±0.21 |
| 6584 [N II] | 0.85 ±0.07 | 0.94 ±0.07 | 0.87 ±0.08 | 0.93 ±0.07 | 0.89 ±0.06 | 0.85 ±0.07 | 0.81 ±0.08 | 0.73 ±0.16 |
| 6717 [S II] | 0.21 ±0.06 | 0.22 ±0.05 | ··· | ··· | 0.19 ±0.06 | 0.35 ±0.05 | 0.42 ±0.07 | ··· |
| 6731 [S II] | 0.20 ±0.06 | 0.41 ±0.05 | ··· | ··· | 0.11 ±0.06 | 0.23 ±0.06 | 0.33 ±0.07 | ··· |
| c(Hβ) | 0.7 ±0.1 | 0.8 ±0.1 | 0.2 ±0.1 | 0.2 ±0.1 | 0.4 ±0.1 | 0.2 ±0.1 | 0.4 ±0.1 | 0.7 ±0.2 |
| EW(Hβ) | 67 | 39 | 12 | 70 | 26 | 78 | 123 | 32 |

TABLE 2b
Dereddened Spectra

| λ Identification | NGC 4651 +014 +007 | NGC 4651 +001 −014 | NGC 4651 −026 +001 | NGC 4651 +048 −023 | NGC 4651 −059 −013 | NGC 4651 −077 −043 | NGC 4651 +131 +021 |
|---|---|---|---|---|---|---|---|
| 3727 [O II] | 1.17 ±0.11 | 1.63 ±0.14 | 2.66 ±0.21 | 3.23 ±0.26 | 2.98 ±0.24 | 3.86 ±0.30 | 2.85 ±0.23 |
| 4101 Hδ | ··· | 0.27 ±0.07 | 0.24 ±0.04 | ··· | 0.24 ±0.04 | 0.36 ±0.03 | 0.25 ±0.05 |
| 4340 Hγ | 0.48 ±0.07 | 0.49 ±0.05 | 0.50 ±0.03 | 0.54 ±0.05 | 0.49 ±0.04 | 0.51 ±0.04 | 0.46 ±0.04 |
| 4861 Hβ | 1.00 ±0.06 | 1.00 ±0.05 | 1.00 ±0.05 | 1.00 ±0.05 | 1.00 ±0.06 | 1.00 ±0.05 | 1.00 ±0.06 |
| 4959 [O III] | ··· | 0.07 ±0.02 | 0.31 ±0.02 | 0.55 ±0.06 | 0.44 ±0.04 | 0.62 ±0.04 | 0.30 ±0.03 |
| 5007 [O III] | 0.10 ±0.04 | 0.17 ±0.03 | 0.89 ±0.05 | 1.23 ±0.08 | 1.25 ±0.07 | 1.76 ±0.09 | 0.99 ±0.06 |
| 6563 Hα | 2.86 ±0.15 | 2.86 ±0.14 | 2.86 ±0.14 | 2.86 ±0.14 | 2.86 ±0.16 | 2.86 ±0.14 | 2.80 ±0.16 |
| 6584 [N II] | 1.17 ±0.07 | 1.25 ±0.08 | 0.74 ±0.05 | 0.61 ±0.07 | 0.71 ±0.07 | 0.54 ±0.04 | 0.59 ±0.08 |
| 6717 [S II] | 0.45 ±0.05 | 0.36 ±0.04 | 0.28 ±0.03 | 0.14 ±0.07 | 0.53 ±0.06 | 0.28 ±0.03 | 0.40 ±0.07 |
| 6731 [S II] | 0.27 ±0.04 | 0.25 ±0.03 | 0.19 ±0.03 | 0.15 ±0.07 | 0.21 ±0.05 | 0.23 ±0.03 | 0.23 ±0.06 |
| c(Hβ) | 0.6 ±0.1 | 0.4 ±0.1 | 0.2 ±0.1 | 0.3 ±0.1 | 0.1 ±0.1 | 0.4 ±0.1 | 0.0 ±0.1 |
| EW(Hβ) | 86 | 32 | 72 | 64 | 170 | 179 | 155 |

TABLE 2c

Dereddened Spectra

| λ Identification | NGC 4654 −068 +033 | NGC 4654 −034 −056 | NGC 4654 −055 +051 | NGC 4654 −042 +035 | NGC 4654 +004 −003 | NGC 4654 +015 −029 | NGC 4654 −019 +013 |
|---|---|---|---|---|---|---|---|
| 3727 [O II] | 2.97 ±0.23 | 3.06 ±0.24 | 3.27 ±0.26 | 2.06 ±0.16 | 0.50 ±0.15 | 1.66 ±0.13 | 1.07 ±0.14 |
| 4101 Hδ | 0.29 ±0.03 | 0.29 ±0.04 | 0.37 ±0.04 | ··· | ··· | 0.28 ±0.03 | ··· |
| 4340 Hγ | 0.54 ±0.03 | 0.54 ±0.04 | 0.48 ±0.04 | 0.43 ±0.04 | ··· | 0.41 ±0.03 | ··· |
| 4861 Hβ | 1.00 ±0.05 | 1.00 ±0.06 | 1.00 ±0.06 | 1.00 ±0.06 | 1.00 ±0.09 | 1.00 ±0.05 | 1.00 ±0.09 |
| 4959 [O III] | 0.29 ±0.02 | 0.48 ±0.04 | 0.53 ±0.03 | ··· | ··· | 0.15 ±0.02 | ··· |
| 5007 [O III] | 0.82 ±0.04 | 1.35 ±0.07 | 1.31 ±0.07 | 0.46 ±0.04 | 0.22 ±0.06 | 0.46 ±0.03 | $\leq$ 0.11 |
| 6563 Hα | 2.86 ±0.14 | 2.86 ±0.14 | 2.86 ±0.15 | 2.86 ±0.15 | 2.86 ±0.17 | 2.84 ±0.14 | 2.86 ±0.16 |
| 6584 [N II] | 0.91 ±0.05 | 0.47 ±0.05 | 0.52 ±0.05 | 0.89 ±0.06 | 1.05 ±0.10 | 0.84 ±0.04 | 0.84 ±0.08 |
| 6717 [S II] | 0.38 ±0.03 | 0.32 ±0.05 | 0.34 ±0.04 | 0.35 ±0.04 | ··· | 0.20 ±0.03 | ··· |
| 6731 [S II] | 0.27 ±0.03 | 0.29 ±0.05 | 0.23 ±0.04 | 0.26 ±0.04 | ··· | 0.12 ±0.04 | ··· |
| c(Hβ) | 0.5 ±0.1 | 0.1 ±0.1 | 0.3 ±0.1 | 0.3 ±0.1 | 0.7 ±0.2 | 0.3 ±0.0 | 0.5 ±0.1 |
| EW(Hβ) | 83 | 100 | 127 | 59 | 10 | 280 | 15 |

TABLE 2d

Dereddened Spectra

| λ Identification | NGC 4689 +016 −018 | NGC 4689 −028 −019 | NGC 4689 +052 +007 | NGC 4689 +014 +010 | NGC 4713 −019 −022 | NGC 4713 +012 −028 | NGC 4713 +005 +014 | NGC 4713 +042 −002 |
|---|---|---|---|---|---|---|---|---|
| 3727 [O II] | 0.48 ±0.10 | 0.65 ±0.08 | 0.95 ±0.10 | 0.38 ±0.07 | 2.86 ±0.22 | 3.36 ±0.26 | 2.38 ±0.19 | 3.80 ±0.30 |
| 4101 Hδ | 0.26 ±0.08 | 0.23 ±0.08 | 0.27 ±0.08 | 0.31 ±0.10 | 0.26 ±0.02 | 0.32 ±0.03 | 0.25 ±0.07 | 0.22 ±0.06 |
| 4340 Hγ | 0.46 ±0.09 | 0.42 ±0.07 | 0.52 ±0.09 | 0.45 ±0.08 | 0.44 ±0.02 | 0.47 ±0.03 | 0.38 ±0.04 | 0.57 ±0.05 |
| 4861 Hβ | 1.00 ±0.06 | 1.00 ±0.07 | 1.00 ±0.08 | 1.00 ±0.08 | 1.00 ±0.05 | 1.00 ±0.05 | 1.00 ±0.05 | 1.00 ±0.05 |
| 4959 [O III] | ··· | ··· | ··· | ··· | 0.59 ±0.03 | 0.44 ±0.06 | 0.29 ±0.03 | 0.70 ±0.05 |
| 5007 [O III] | $\leq$ 0.07 | 0.10 ±0.05 | $\leq$ 0.10 | $\leq$ 0.09 | 1.77 ±0.09 | 1.34 ±0.07 | 0.82 ±0.04 | 2.05 ±0.10 |
| 6563 Hα | 2.86 ±0.16 | 2.86 ±0.16 | 2.86 ±0.16 | 2.86 ±0.15 | 2.86 ±0.14 | 2.86 ±0.14 | 2.86 ±0.14 | 2.86 ±0.14 |
| 6584 [N II] | 0.71 ±0.08 | 0.90 ±0.08 | 0.65 ±0.08 | 0.63 ±0.06 | 0.44 ±0.02 | 0.22 ±0.03 | 0.47 ±0.05 | 0.27 ±0.05 |
| 6717 [S II] | 0.31 ±0.09 | 0.26 ±0.07 | ··· | 0.27 ±0.08 | 0.22 ±0.02 | 0.42 ±0.04 | 0.37 ±0.05 | 0.31 ±0.06 |
| 6731 [S II] | ··· | ··· | ··· | ··· | 0.17 ±0.02 | 0.17 ±0.03 | 0.20 ±0.04 | 0.20 ±0.06 |
| c(Hβ) | 0.3 ±0.1 | 0.4 ±0.1 | 0.1 ±0.1 | 0.3 ±0.1 | 0.3 ±0.1 | 0.1 ±0.1 | 0.2 ±0.0 | 0.3 ±0.1 |
| EW(Hβ) | 81 | 67 | 38 | 26 | 182 | 80 | 31 | 100 |

TABLE 3
Virgo Sample H II Region Abundances

| HII Region | $R/R_E$ | $\log(R_{23})$ | 12 + log(O/H) | Reference |
|---|---|---|---|---|
| | | NGC 4254 | | |
| +013 +006 | 0.28 | −0.33±0.08 | 9.36±0.02 (0.18) | MRS |
| −011 −018 | 0.41 | −0.37±0.10 | 9.36±0.02 (0.18) | HPC |
| +031 −011 | 0.69 | −0.03±0.07 | 9.27±0.02 (0.16) | HPC |
| −031 −022 | 0.73 | −0.14±0.05 | 9.31±0.01 (0.17) | HPC |
| −005 +042 | 0.90 | 0.09±0.05 | 9.23±0.02 (0.15) | MRS |
| +006 −051 | 1.09 | 0.09±0.05 | 9.23±0.02 (0.15) | HPC |
| −007 −066 | 1.37 | 0.22±0.04 | 9.18±0.02 (0.14) | HPC |
| +055 −042 | 1.48 | 0.30±0.06 | 9.14±0.03 (0.13) | MRS |
| +080 −016 | 1.67 | 0.39±0.05 | 9.09±0.03 (0.12) | SSK |
| −039 −076 | 1.69 | 0.29±0.07 | 9.14±0.04 (0.13) | HPC |
| −047 −075 | 1.73 | 0.30±0.04 | 9.14±0.02 (0.13) | MRS |
| +093 +039 | 1.94 | 0.49±0.06 | 9.01±0.05 (0.13) | MRS |
| −039 −092 | 1.99 | 0.49±0.03 | 9.01±0.03 (0.13) | HPC |
| +102 +015 | 2.03 | 0.52±0.05 | 8.99±0.05 (0.14) | MRS |
| +025 +100 | 2.09 | 0.57±0.05 | 8.94±0.05 (0.14) | HPC |
| +106 +019 | 2.11 | 0.48±0.04 | 9.02±0.03 (0.13) | SSK |
| −117 −002 | 2.33 | 0.34±0.05 | 9.12±0.03 (0.13) | SSK |
| +077 +101 | 2.47 | 0.64±0.02 | 8.86±0.02 (0.15) | HPC |
| +090 +102 | 2.63 | 0.59±0.06 | 8.92±0.07 (0.14) | HPC |
| | | NGC 4303 | | |
| +021 −007 | 0.37 | 0.00±0.14 | 9.27±0.05 (0.16) | SSK |
| +021 −007 | 0.37 | 0.14±0.17 | 9.21±0.07 (0.15) | HPLC |
| −001 +045 | 0.68 | 0.01±0.06 | 9.26±0.02 (0.16) | SSK |
| +010 −044 | 0.69 | 0.24±0.05 | 9.17±0.02 (0.13) | HPLC |
| −013 −044 | 0.69 | 0.09±0.05 | 9.23±0.02 (0.15) | SSK |
| −013 −044 | 0.69 | 0.17±0.04 | 9.20±0.02 (0.14) | HPLC |
| −025 −042 | 0.75 | 0.29±0.05 | 9.14±0.03 (0.13) | HPLC |
| +045 −008 | 0.76 | 0.50±0.18 | 9.00±0.15 (0.13) | HPLC |
| −014 +048 | 0.76 | 0.10±0.05 | 9.25±0.02 (0.15) | SSK |
| +046 +006 | 0.76 | −0.03±0.16 | 9.27±0.05 (0.16) | SSK |
| +032 −040 | 0.81 | 0.39±0.05 | 9.09±0.03 (0.12) | SSK |
| +032 −040 | 0.81 | 0.49±0.10 | 9.01±0.08 (0.13) | HPLC |
| +043 −026 | 0.82 | 0.33±0.21 | 9.12±0.12 (0.13) | HPLC |
| +016 −057 | 0.90 | 0.30±0.19 | 9.14±0.11 (0.13) | HPLC |
| +022 +067 | 1.06 | 0.04±0.06 | 9.25±0.02 (0.16) | SSK |
| +005 −073 | 1.10 | 0.23±0.05 | 9.17±0.03 (0.14) | SSK |
| +005 −073 | 1.10 | 0.54±0.28 | 8.97±0.28 (0.14) | HPLC |
| −008 −089 | 1.34 | 0.58±0.27 | 8.93±0.30 (0.14) | HPLC |
| −049 −094 | 1.61 | 0.68±0.04 | 8.81±0.05 (0.15) | SSK |
| −110 +075 | 2.17 | 0.76±0.04 | 8.70±0.07 (0.14) | SSK |
| −110 +075 | 2.17 | 0.74±0.05 | 8.73±0.08 (0.15) | HPLC |
| −070 +140 | 2.43 | 0.90±0.05 | 8.44±0.11 (0.08) | HPLC |

TABLE 3—*Continued*

| HII Region | $R/R_E$ | $\log(R_{23})$ | 12 + log(O/H) | Reference |
|---|---|---|---|---|
| | | NGC 4321 | | |
| +008 −004 | 0.11 | −0.06±0.10 | 9.28±0.03 (0.17) | MRS |
| −051 +009 | 0.68 | −0.08±0.06 | 9.29±0.02 (0.17) | MRS |
| −001 −066 | 0.82 | 0.10±0.07 | 9.26±0.03 (0.15) | SSK |
| +032 −074 | 0.97 | −0.11±0.06 | 9.30±0.02 (0.17) | MRS |
| +013 +102 | 1.29 | 0.14±0.06 | 9.21±0.03 (0.15) | SSK |
| −114 +010 | 1.53 | 0.21±0.05 | 9.18±0.02 (0.14) | MRS |
| −032 +147 | 1.82 | 0.54±0.05 | 8.97±0.05 (0.14) | SSK |
| −131 −027 | 1.83 | 0.40±0.05 | 9.08±0.04 (0.12) | SSK |
| +029 +146 | 1.89 | 0.35±0.06 | 9.11±0.04 (0.12) | SSK |
| +034 +145 | 1.90 | 0.27±0.05 | 9.16±0.03 (0.13) | MRS |
| | | NGC 4501 | | |
| +018 +012 | 0.48 | −0.66±0.12 | 9.39±0.03 (0.18) | SKSZ |
| −043 −005 | 0.86 | 0.00±0.05 | 9.27±0.02 (0.16) | SKSZ |
| −027 +062 | 0.88 | −0.57±0.10 | 9.39±0.01 (0.17) | SKSZ |
| −068 +093 | 1.38 | −0.20±0.05 | 9.32±0.02 (0.17) | SKSZ |
| −094 +049 | 1.50 | 0.30±0.03 | 9.14±0.02 (0.13) | SKSZ |
| | | NGC 4571 | | |
| −029 −011 | 0.60 | −0.21±0.18 | 9.33±0.06 (0.17) | SKSZ |
| −010 +050 | 1.09 | −0.05±0.10 | 9.28±0.04 (0.17) | SSK |
| +055 +018 | 1.12 | 0.09±0.05 | 9.23±0.02 (0.15) | SKSZ |
| −017 −062 | 1.30 | −0.20±0.05 | 9.32±0.01 (0.17) | SKSZ |
| | | NGC 4651 | | |
| +014 +007 | 0.39 | 0.11±0.04 | 9.22±0.02 (0.15) | SKSZ |
| +001 −014 | 0.47 | 0.27±0.03 | 9.15±0.02 (0.13) | SKSZ |
| −026 +001 | 0.65 | 0.59±0.02 | 8.92±0.03 (0.14) | SKSZ |
| +048 −023 | 1.51 | 0.70±0.02 | 8.78±0.03 (0.15) | SKSZ |
| +059 −013 | 1.59 | 0.67±0.05 | 8.82±0.03 (0.15) | SKSZ |
| −077 −043 | 2.20 | 0.80±0.02 | 8.64±0.04 (0.13) | SKSZ |
| +131 +021 | 3.21 | 0.62±0.02 | 8.89±0.03 (0.15) | SKSZ |
| | | NGC 4654 | | |
| +004 −003 | 0.08 | −0.10±0.08 | 9.30±0.03 (0.17) | SKSZ |
| −019 +013 | 0.35 | 0.03±0.05 | 9.26±0.02 (0.16) | SKSZ |
| +015 −029 | 0.56 | 0.36±0.03 | 9.10±0.02 (0.12) | SKSZ |
| −042 +035 | 0.82 | 0.42±0.03 | 9.06±0.02 (0.12) | SKSZ |
| −055 +051 | 1.13 | 0.71±0.02 | 8.77±0.03 (0.15) | SKSZ |
| −068 +033 | 1.18 | 0.61±0.03 | 8.89±0.03 (0.15) | SKSZ |
| −034 −056 | 1.59 | 0.69±0.02 | 8.80±0.03 (0.15) | SKSZ |
| | | NGC 4689 | | |
| +014 +010 | 0.29 | −0.42±0.07 | 9.37±0.01 (0.18) | SKSZ |
| +016 −018 | 0.37 | −0.31±0.08 | 9.35±0.02 (0.18) | SKSZ |
| −028 −019 | 0.58 | −0.11±0.05 | 9.30±0.02 (0.17) | SKSZ |
| −033 −011 | 0.60 | 0.31±0.09 | 9.14±0.06 (0.13) | SSK |
| +052 −007 | 0.91 | −0.02±0.04 | 9.27±0.02 (0.16) | SKSZ |
| | | NGC 4713 | | |
| +005 +014 | 0.82 | 0.54±0.02 | 8.96±0.02 (0.14) | SKSZ |
| −019 −022 | 1.53 | 0.72±0.02 | 8.76±0.03 (0.15) | SKSZ |
| +042 −002 | 1.61 | 0.82±0.02 | 8.60±0.04 (0.12) | SKSZ |
| +012 −028 | 1.63 | 0.71±0.02 | 8.77±0.03 (0.15) | SKSZ |

TABLE 4

CHARACTERISTIC ABUNDANCES AND ABUNDANCE GRADIENTS

| Name | 12 + log (O/H) at 0.4 $R_O$ | Gradient ($dex/R_O$) |
|---|---|---|
| NGC 4254 | 9.18±0.02 | −0.63±0.17 |
| NGC 4303 | 9.09±0.02 | −1.13±0.13 |
| NGC 4321 | 9.23±0.02 | −0.39±0.22 |
| NGC 4501 | 9.32±0.05 | −0.53±0.48 |
| NGC 4571 | 9.24±0.02 | −0.13±0.69 |
| NGC 4651 | 8.99±0.06 | −0.42±0.17 |
| NGC 4654 | 9.01±0.03 | −0.84±0.27 |
| NGC 4689 | 9.28±0.02 | −0.37±0.70 |
| NGC 4713 | 8.84±0.03 | −1.08±0.65 |

TABLE 5

COMPARISON OF AVERAGE CHARACTERISTIC PROPERTIES

| Property | Virgo HI Deficient | Virgo Intermediate | Virgo HI Normal | Field |
|---|---|---|---|---|
| 12 +log (O/H) | 9.28 | 9.14 | 8.97 | ... |
| EW (H$\alpha$) (Å) | 9.7 | 22 | 37 | 31 |
| B-V | 0.58 | 0.57 | 0.49 | 0.51 |
| $\tau_{\rm R}$ ($10^9$ yrs) | 3.77 | 2.80 | 2.40 | 4.59 |
| $\tau_{\rm R,d}$ ($10^9$ yrs) | 3.77 | 2.67 | 1.27 | 3.11 |

TABLE 6

RESULTS OF MODEL CALCULATIONS

| Model | SHN | SHC | SLN | SLC | TLN | TLC | THN | THC |
|---|---|---|---|---|---|---|---|---|
| infall rate | 0.2 | 0.2 | 0.05 | 0.05 | 0.25 | 0.25 | 0.65 | 0.65 |
| $\tau_s$ | 2.5 | 2.5 | 2.5 | 2.5 | ... | ... | ... | ... |
| $\tau_f$ | ... | ... | ... | ... | 50.0 | 50.0 | 8.0 | 8.0 |
| $M_g$ | 0.90 | 0.49 | 0.32 | 0.22 | 1.0 | 1.0 | 1.0 | 1.0 |
| $M_s$, gross | 3.50 | 3.20 | 2.02 | 1.96 | 3.80 | 2.90 | 6.40 | 5.40 |
| $M_s$, net | 2.10 | 1.91 | 1.18 | 1.13 | 2.30 | 1.63 | 3.70 | 3.00 |
| $\mu$ | 0.30 | 0.20 | 0.21 | 0.17 | 0.31 | 0.38 | 0.21 | 0.25 |
| Z | 1.49 | 2.30 | 2.02 | 2.69 | 1.49 | 1.48 | 1.76 | 1.91 |
| Z (simple) | 2.02 | 2.70 | 2.67 | 3.12 | 1.95 | 1.71 | 2.65 | 2.47 |
| Z (Larson) | 1.64 | 1.67 | 1.67 | 1.72 | 1.48 | 1.42 | 1.67 | 1.69 |
| SFR | 0.36 | 0.20 | 0.13 | 0.09 | 0.38 | 0.05 | 0.40 | 0.08 |
| $M_g$/SFR | 2.50 | 2.50 | 2.50 | 2.50 | 2.63 | 21.3 | 2.50 | 12.3 |
| $b$ | 1.02 | 0.61 | 0.64 | 0.45 | 1.01 | 0.16 | 0.63 | 0.15 |

This manuscript was prepared with the AAS LaTeX macros v3.0.

Fig. 1.— A map of the positions of the nine Virgo spiral galaxies studied here, and the structures identified by Binggeli, Tammann, & Sandage (1987). The points are coded to represent the three subgroups into which the sample has been divided (see text). The filled circles represent the H I deficient galaxies, the filled squares represent the intermediate galaxies, and the filled triangles represent the galaxies with normal H I disks.

Fig. 2.— A comparison of the neutral hydrogen to optical diameter ratios ($D_H/D_O$) for the field galaxy sample of Warmels (1988) and the Virgo spirals studied here. The connected line represents the average value as a function of T-type for the field sample of Warmels. The error bars represent the r.m.s. standard deviation for each T-type sample. The solid symbols represent the nine Virgo galaxies studied here, and are shown with T-types taken from the RC3. The symbol coding is identical to that in Figure 1.

Fig. 3.— A comparison of the radial H I profiles of the Virgo spiral galaxies in our sample with the average radial H I profile for the field galaxies of corresponding Hubble type as derived by Cayatte et al. (1994). The galaxies are ordered in increasing $D_H/D_O$ as taken from Figure 2. The individual points represent the radially averaged H I column densities, while the solid lines represent the average radial H I profile for the corresponding Hubble type. The filled circles represent the WSRT observations of Warmels (1988), while the filled triangles represent the VLA observations of Cayatte et al. (1990), and, in the case of NGC 4571, van der Hulst et al. (1987). There is good agreement between the WSRT and VLA observations as noted by Cayatte et al. (1990).

Fig. 4.— Oxygen abundances from H II regions are plotted versus galactic radius for nine Virgo spiral galaxies. The order of the galaxies is the same as in Figure 3. The oxygen abundances are determined from measurements of [O II] and [O III] in individual H II regions and calibrated empirically as described in the text. The radial positions are normalized to the effective optical radius as defined in the RC3. The filled circles are taken from MRS, SSK, and the new observations presented here. The filled triangles are taken from HPC and HPLC. Note the good agreement between the abundance measurements from the different studies.

Fig. 5.— (a) A plot of mean O/H as a function of $D_H/D_O$ for the Virgo spirals. (b) Oxygen abundance gradient versus $D_H/D_O$. The symbol coding is identical to that in Figure 1.

Fig. 6.— Mean O/H as a function of (a) absolute blue magnitude, (b) maximum rotation curve velocity, and (c) Hubble type for the Virgo spirals and the non-barred spirals of the ZKH sample. The open circles correspond to the field sample of ZKH, while the points for the Virgo spirals have been coded as in Figure 1.

Fig. 7.— $M_B$ and $V_C$ versus T-type for the ZKH and Virgo spiral samples. Coding as in Figure 6.

Fig. 8.— A comparison of the H II region oxygen abundance trends for the H I deficient Virgo cluster galaxies NGC 4571 (T = 6), NGC 4689 (T = 4), and NGC 4501 (T = 3) with field galaxies of comparable Hubble type (taken from ZKH). The H I deficient galaxies are shown as bold lines.

Fig. 9.— A comparison of the oxygen abundances for the individual H II regions in the Virgo cluster galaxies. Top panel – The oxygen abundances are plotted versus the galactocentric radius normalized by the effective radius of the disk. Middle panel – The oxygen abundances are plotted versus the V surface brightness of the disk. Lower panel – The oxygen abundances are plotted versus the radially averaged gas mass fraction (described in text). Filled circles represent HI deficient Virgo galaxies. Open squares represent intermediate cases and filled triangles represent Virgo galaxies with normal HI contents.

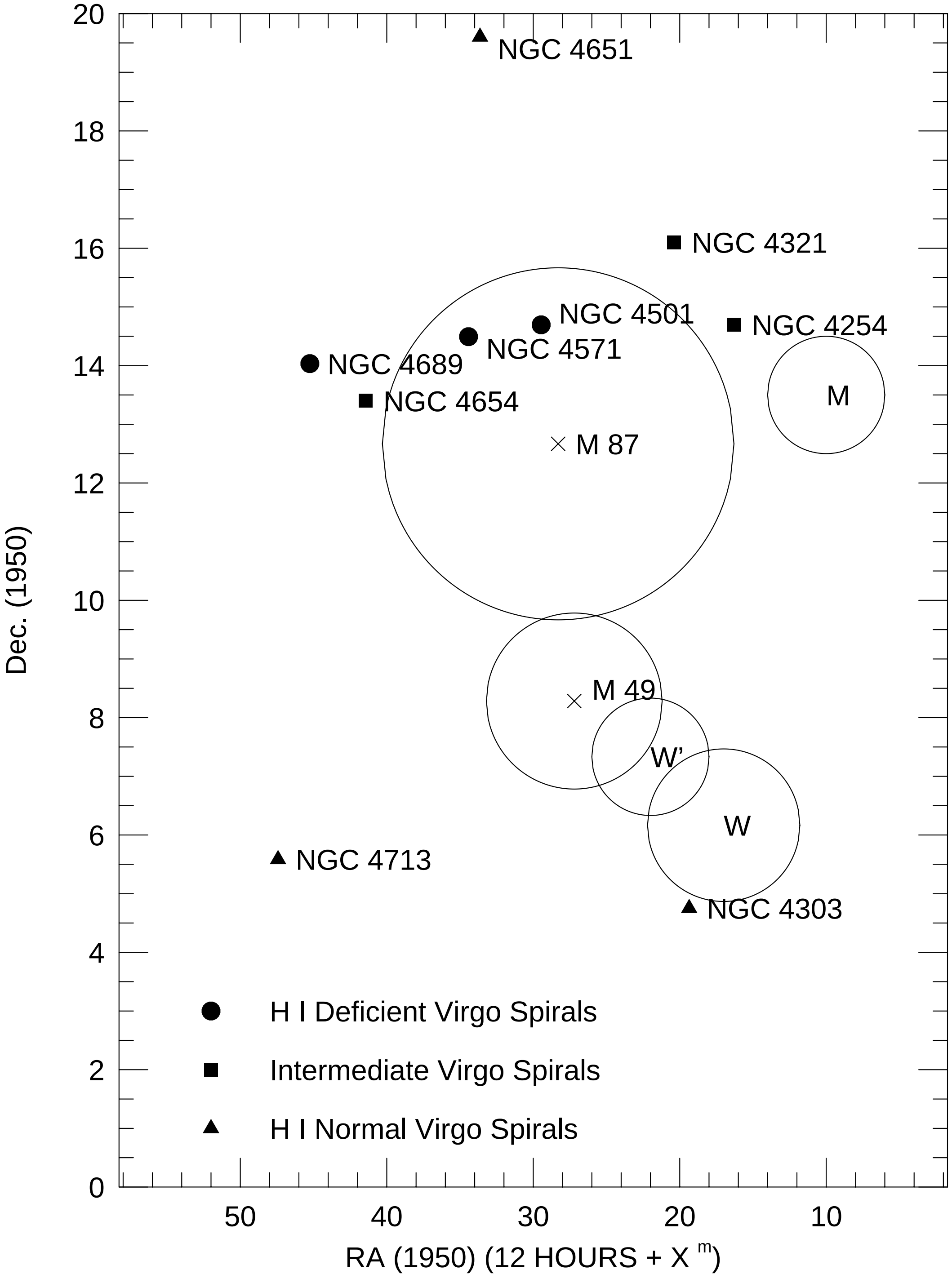

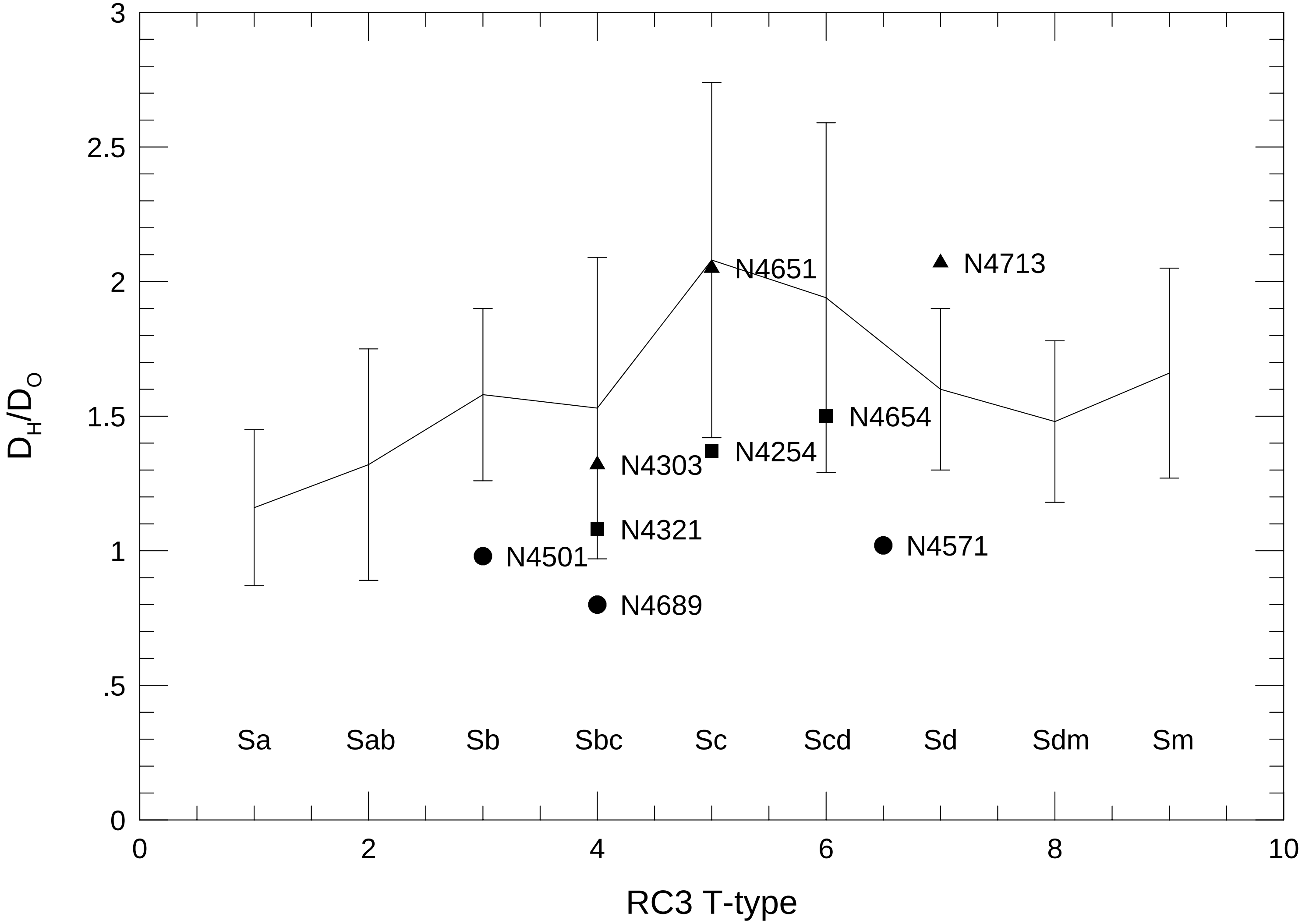

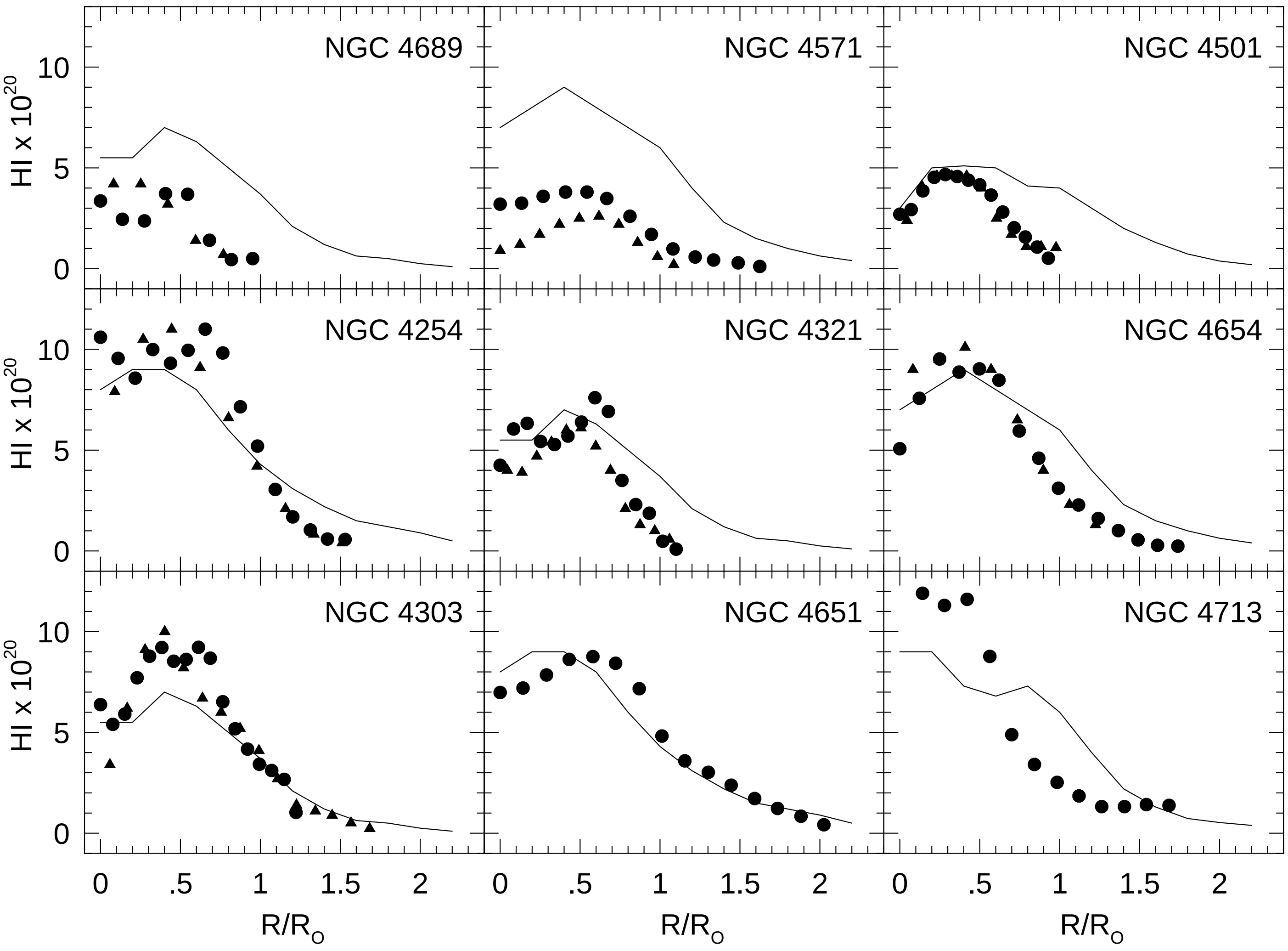

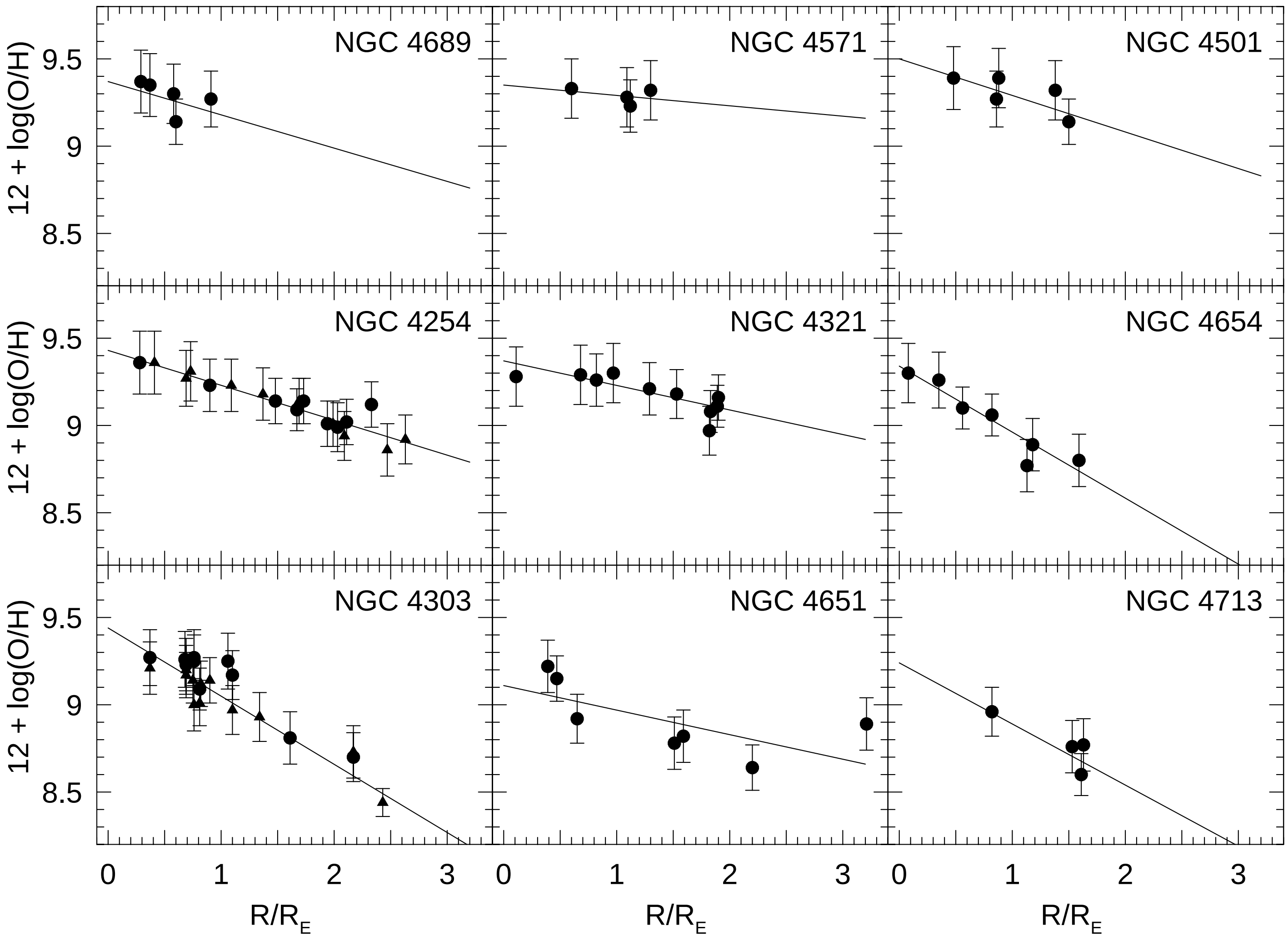

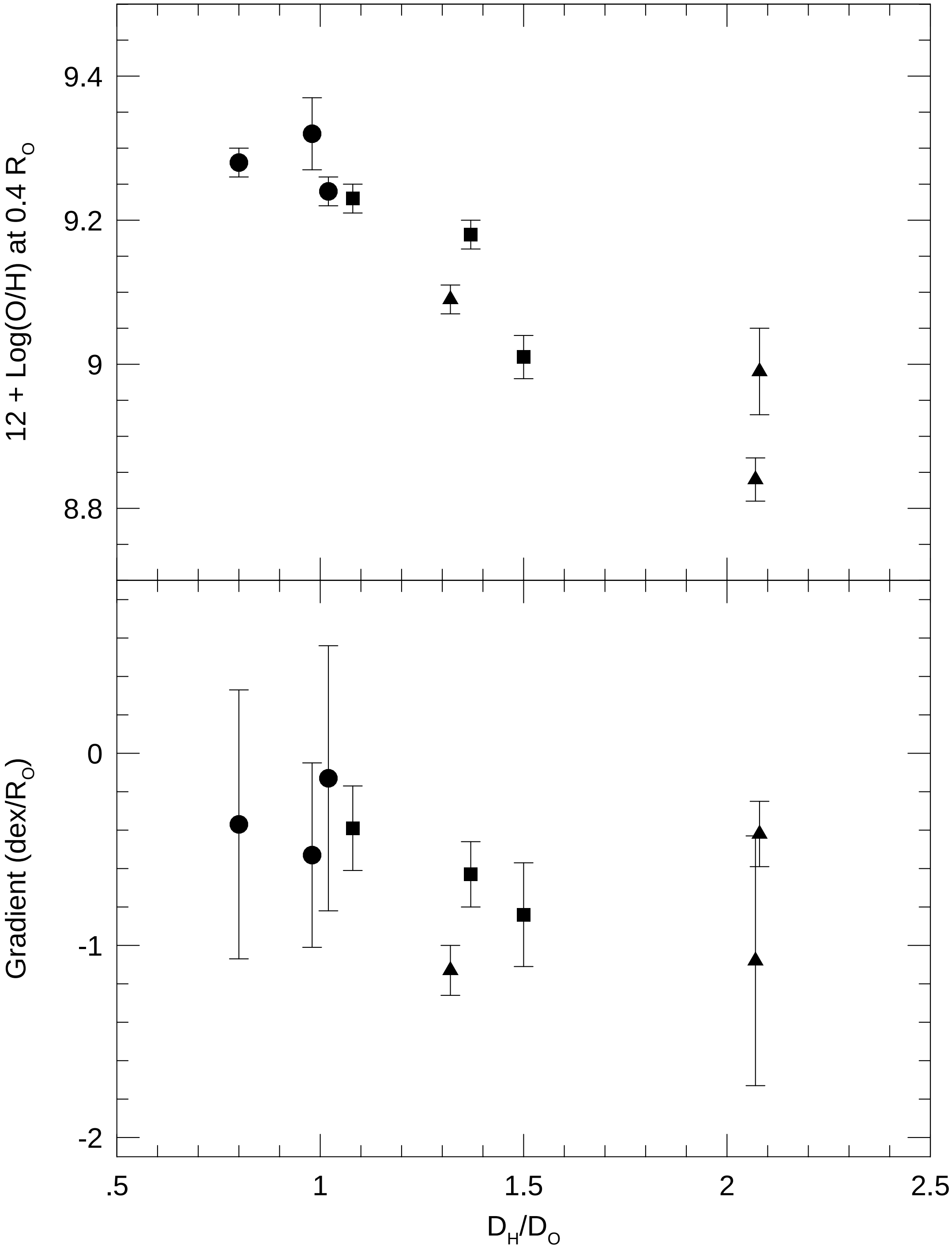

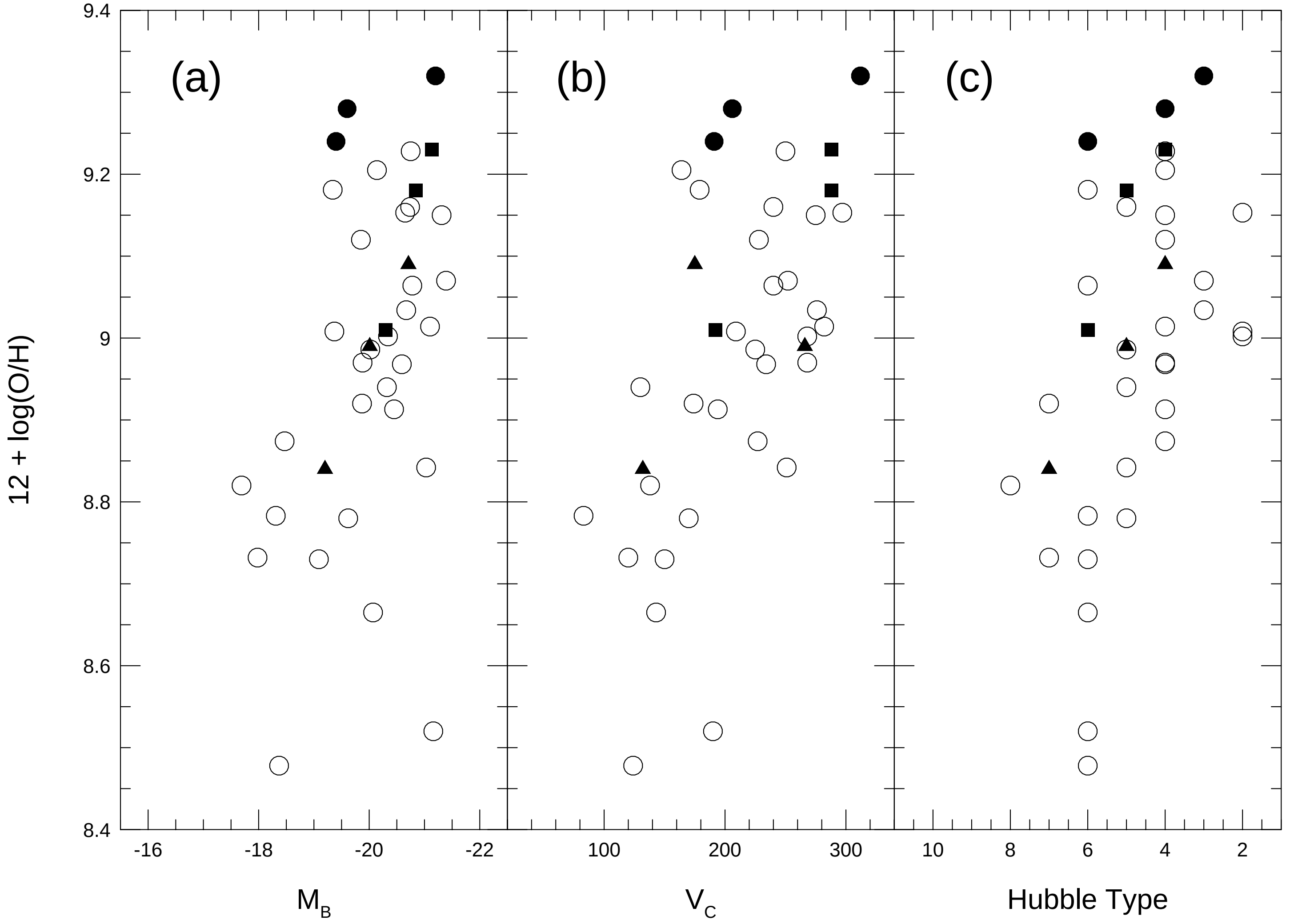

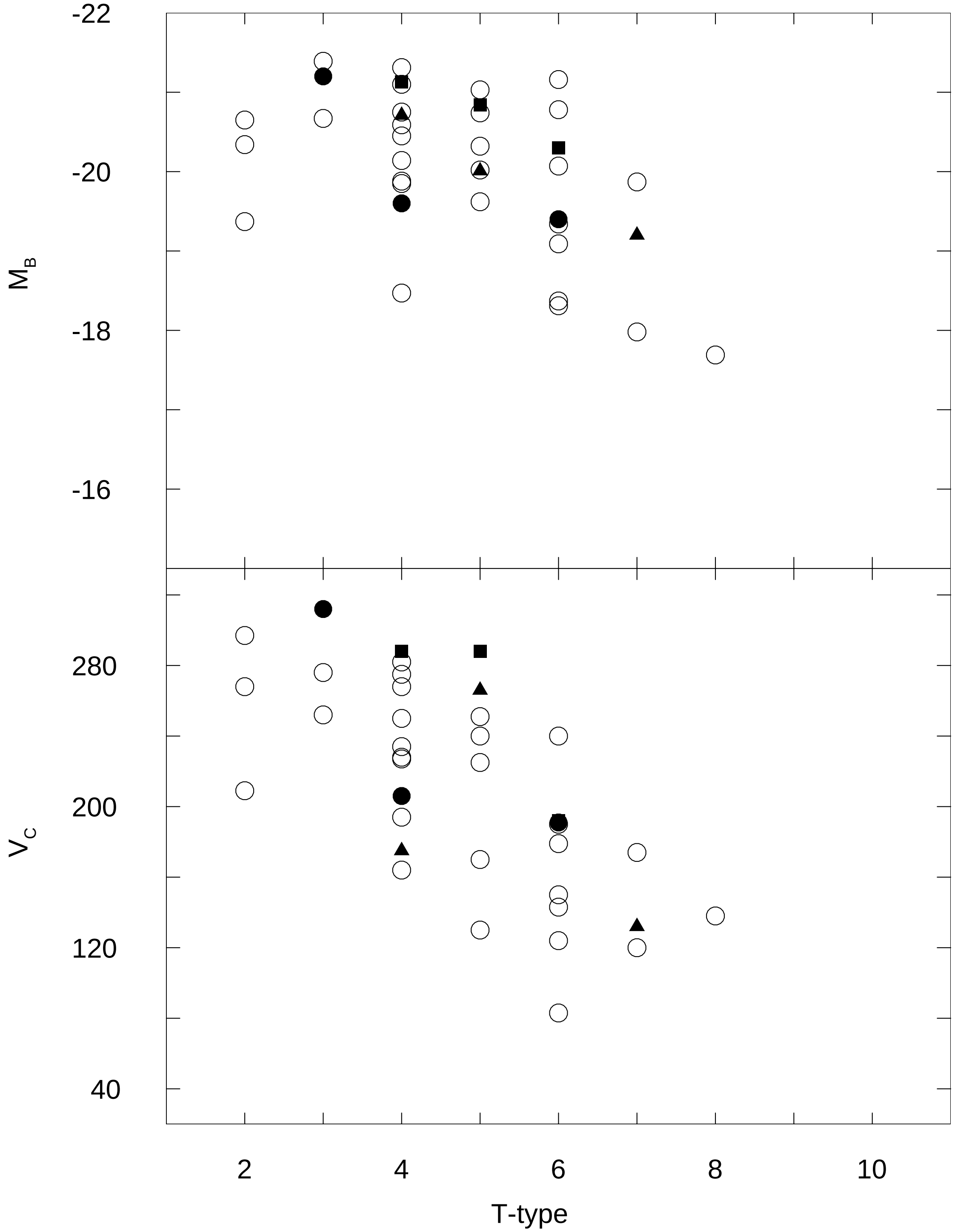

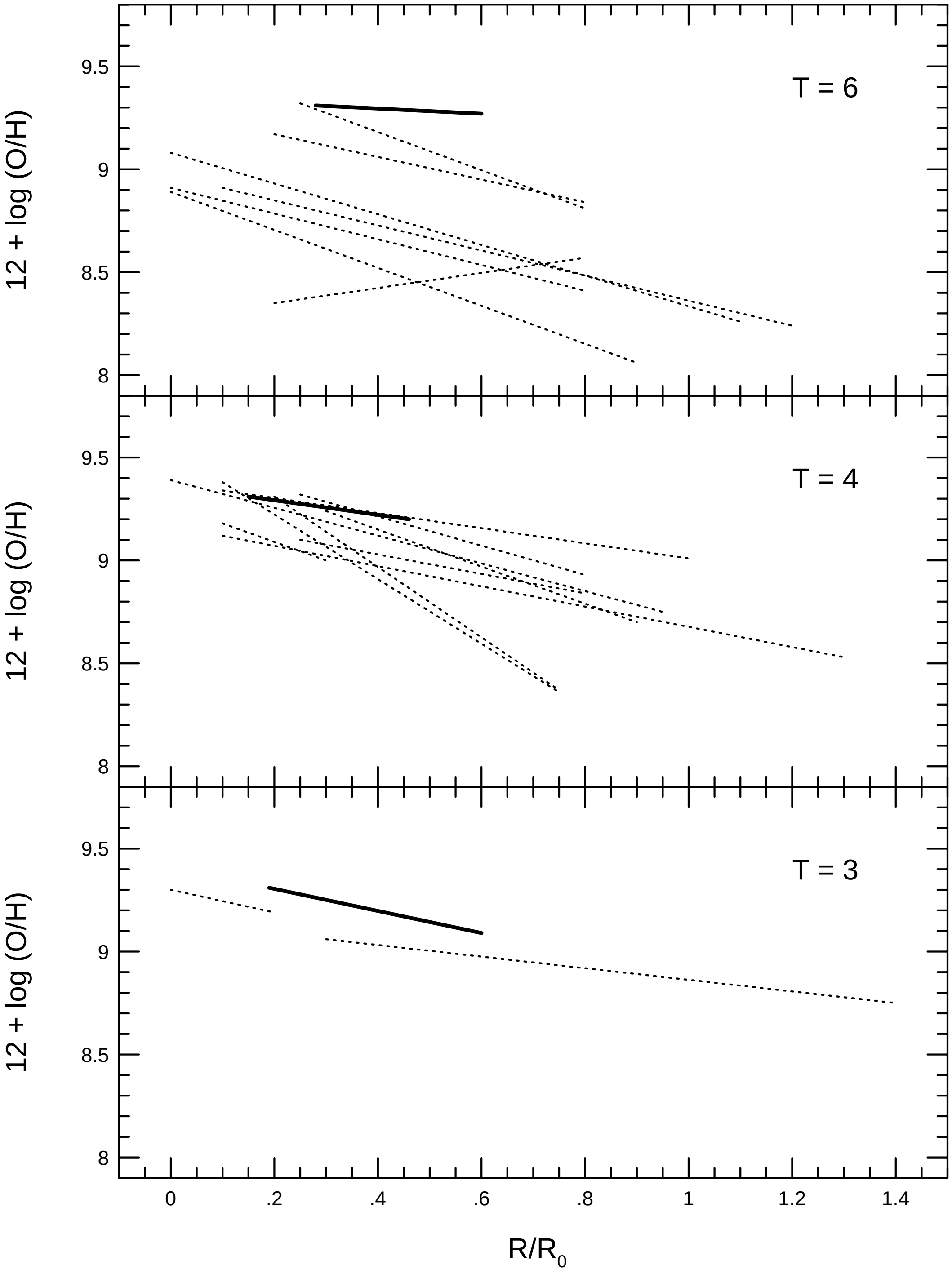

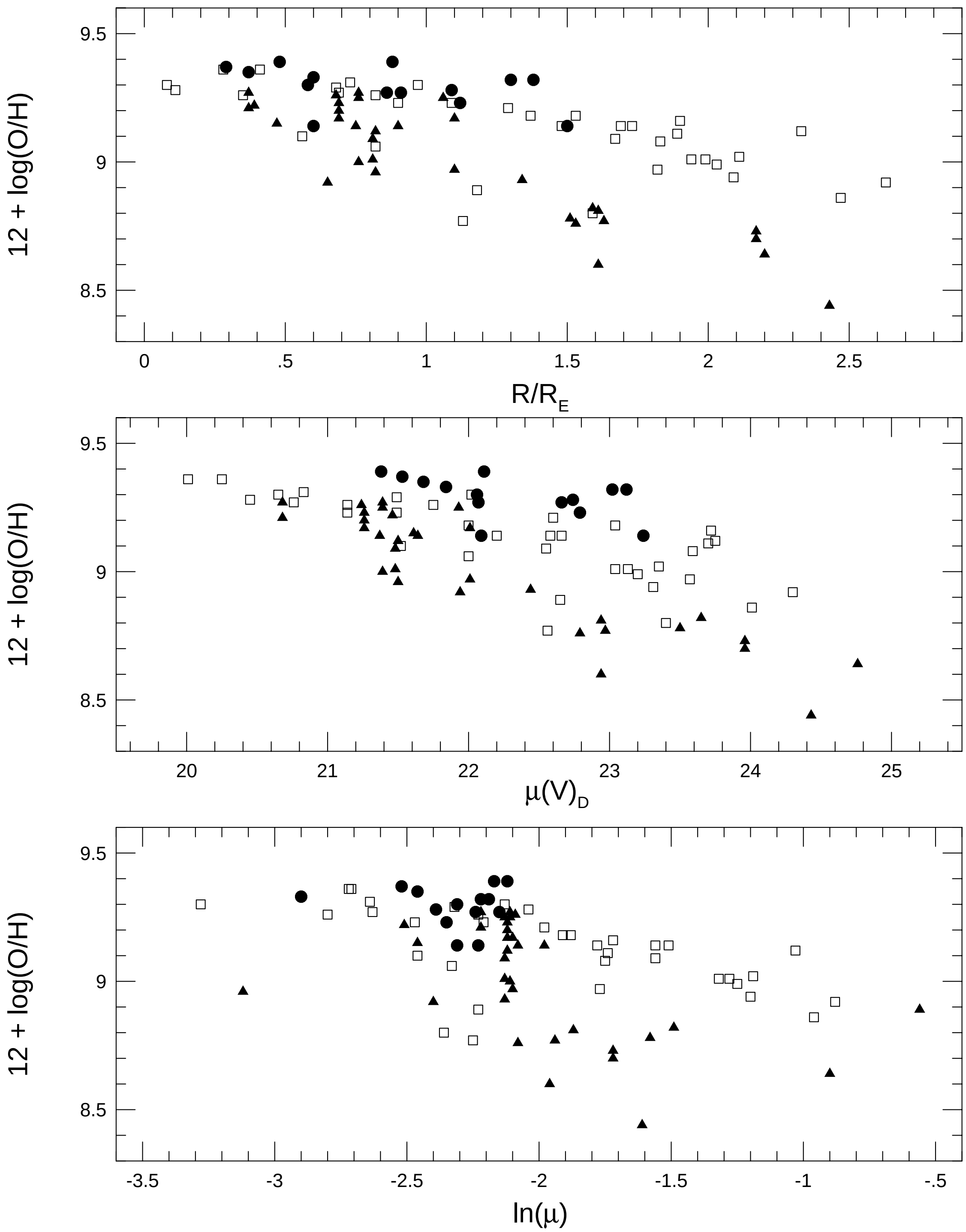